\documentclass[twocolumn,aps,showpacs,prb,tightenlines,amsmath,amssymb]{revtex4}
\usepackage{graphicx}    
\usepackage{amssymb}  
\usepackage{dcolumn}
\usepackage{amsmath}     
\usepackage{bm}                                          
\usepackage{colordvi}                   

\begin{document}             
\title{Multivalley spin relaxation in
  $n$-type bulk GaAs in the presence of high electric fields}
\author{H. Tong}
\affiliation{Hefei National Laboratory for Physical Sciences at Microscale and
  Department of Physics, University of Science and Technology of China, Hefei,
  Anhui, 230026, China} 
\author{M. W. Wu}
\thanks{Author to  whom correspondence should be addressed}
\email{mwwu@ustc.edu.cn.}
\affiliation{Hefei National Laboratory for Physical Sciences at Microscale and
  Department of Physics, University of Science and Technology of China, Hefei,
  Anhui, 230026, China} 
\date{\today}

\begin{abstract}
Multivalley spin relaxation in $n$-type bulk GaAs in the presence of high
electric field is investigated from the microscopic kinetic spin Bloch equation 
approach with the $\Gamma$ and $L$ valleys included. We show that the spin
relaxation time decreases monotonically with the electric field, which differs
from the two-dimensional case and is recognized due to to the cubic form of the
Dresselhauss spin-orbit coupling of the $\Gamma$ valley in bulk. In addition to
the direct modulation of the spin relaxation time, the electric field also
strongly influences the density and temperature dependences of
the spin relaxation. 
In contrast to the monotonic decrease with increasing lattice
temperature in the field-free condition, the
spin relaxation time is shown to decrease more slowly under the influence of the
electric field and even to increase monotonically 
in the case with small electron density and high electric field. We even 
predict a peak in weakly doped samples under
moderate electric field  due to the
anomalous lattice-temperature dependence of the
hot-electron temperature. As 
for the $L$ valleys, we show that instead of playing the role of a ``drain''of
the total spin polarization as in quantum well systems, in bulk they serve as
a ``momentum damping area'', which prevents electrons from  drifting to
higher momentum states. This tends to suppress the inhomogeneous broadening and
hence leads to an increase of the spin relaxation time.   

\end{abstract}



\maketitle

\section{INTRODUCTION}
Semiconductor spintronics, which aims at incorporating the spin degree of
freedom in electronics, has attracted much attention in the past  
decades.\cite{Meier,Wolf,Spin1,Spin11,Spin12,Spin13,Wu3,Spin2} One of the
main challenges in realizing spintronic devices lies in the control of spin
lifetime. Many investigations have been devoted in bulk III-V
semiconductors.\cite{Seymour,Zerr,Kikkawa,Barry,Wu4,Dzhioev,Murdin,Hohage,Litvinenko,Oertel,Jiang1,Litvinenko1,vanderWal,Buss,Bob2,Bob,Cundiff,Shen1,Ma1,Zhu1,Romer,Intronati}   
Various factors on spin relaxation, such as carrier and impurity 
densities,\cite{Seymour,Kikkawa,Wu4,Dzhioev,Murdin,Oertel,Jiang1,Buss,Cundiff,Shen1,Bob,Ma1}
temperature,\cite{Zerr,Murdin,Hohage,Litvinenko,Oertel,Bob2,Zhu1,Jiang1} 
mobility,\cite{Litvinenko} initial spin polarization\cite{Jiang1} and 
electric\cite{Barry,Jiang1} and magnetic fields,\cite{Kikkawa,Dzhioev,Buss,Bob2,Wu4} etc, have
been extensively studied. Remarkably, spin relaxation time (SRT) as long as
130~ns is observed in $n$-type GaAs\cite{Kikkawa,Dzhioev} and the SRT varying by
more than three orders of magnitude with temperature or density is
reported.\cite{Kikkawa} Very recently, Jiang and Wu performed a systematic
investigation on the  electron spin relaxation in bulk III-V semiconductors in
metallic regime from the fully microscopic kinetic spin Bloch equation (KSBE)
approach.\cite{Jiang1} The D'yakonov-Perel' (DP) mechanism\cite{DP1} is
demonstrated to be dominant in $n$-type III-V semiconductors, even in narrow
band ones, and many features in contrast to the 
previous understandings in the 
literature are reported.\cite{Jiang1} Some of their predictions have been soon
realized in the subsequent experimental
investigations.\cite{Litvinenko1,vanderWal,Buss,Bob2,Bob,Cundiff,Shen1,Ma1,Zhu1}
Moreover, the very recent works\cite{Romer,Intronati}  on the spin relaxation
near the metal-to-insulator transition have extended the investigation in that
regime back in 2002.\cite{Dzhioev}  

It is noted that works discussed above mainly concern electron spin relaxation
in the absence of or with relatively low electric fields. In
Ref.~\onlinecite{Jiang1}, an electric field applied up to  2~kV/cm is also
considered and the SRT is found to be effectively
manipulated, especially for the low temperature case, where the SRT is
suppressed down to 1/10 of its 
original value with a small electric field up to 0.05~kV/cm. Clear hot-electron
effect is shown and the different field dependence of the SRT compared to that
in quantum wells is addressed. Nevertheless, since most current electronic devices are
performed in the high-field conditions, new features of the SRT are expected
when electrons are drifted to high valleys.\cite{Kratzer} The different
spin-orbit couplings (SOCs),
momentum relaxations and effective masses in different valleys should have   
pronounced effects on spin dynamics. In fact, investigations on how
the spin relaxation is affected by high electric field have been carried out in
quantum well systems. The hot-electron  effect\cite{Weng1} and the multi-subband
effect\cite{Weng3} on spin dephasing in $n$-type GaAs quantum wells have been
studied. Different field dependences of SRT are observed for different
temperatures, well widths and initial spin polarizations.  Moreover, the
multivalley spin relaxation  under high in-plane electric fields has been
investigated in $n$-type GaAs quantum wells by  taking into account the $\Gamma$
and $L$ valleys.\cite{Zhang1} It is predicted that although the SOC in the $L$
valleys is much larger than that of the $\Gamma$ valley,\cite{Fu}
 the spin polarization in the $L$ valleys shares the same damping rate as that in
the $\Gamma$ valley. The $L$ valleys are pointed out to play the role of a
``drain'' of the total spin polarization due to the large SOC therein. 
Nonmonotonic dependence of the SRT on electric field has been
reported, while the spin Gunn effect,\cite{Flatte} the spontaneous
spin-polarization  generation in the high--electric-field charge Gunn  
region, is pointed out to hardly realize in GaAs
quantum wells.  Despite these works in two-dimensional structures, a detailed
fully microscopic 
study in bulk system under high electric field is still absent.  How the
temperature and electron density dependences of the SRT are affected by the high
electric field is still unclear and the feasibility of the spin Gunn effect
predicted in bulk\cite{Flatte} needs to be checked. Because of the cubic form of
the Dresselhauss  SOC in the $\Gamma$ valley as well as the absence of quantum
confinement in bulk, different behaviors in spin relaxation are expected. 

In this work, we apply the KSBEs to investigate the electron spin relaxation in
$n$-type bulk GaAs in the presence of high electric fields with the $\Gamma$ and $L$
valleys included. The electric field dependence of the SRT is calculated and
found to be monotonic, in contrast with the results in
quantum wells.\cite{Zhang1} We attribute this to the strong
enhancement of inhomogeneous broadening\cite{Wu3,Wu1,Wu2} resulting from the
cubic form of the $\Gamma$-valley Dresselhaus SOC in bulk. The electric
field is also shown to effectively change the density and the temperature
dependences of the SRT. Remarkably, a peak is predicted in the temperature
dependence of the SRT for relative low electron density under a moderate   
electric field. The underlying physics is analyzed. As for the $L$
valleys, we show that in contrast to the two-dimensional system where the $L$
valleys play the role of a ``drain'' of the total spin polarization,
unexpectedly in bulk their effect on the total spin dynamics is weak and the
total spin   relaxation is mainly determined by the $\Gamma$ valley. It is shown
that even with much larger SOC in the $L$ valleys, the inclusion of the 
$L$ valleys results in a {\em longer} rather than a shorter SRT. We find that  
the $L$ valleys serve as a ``momentum damping area'' where electrons are hardly to
drift to higher momentum states due their large effective mass. This tends to
suppress the inhomogeneous broadening and hence leads to the increase of the SRT.  

This paper is organized as follows: In Sec.~II, we set up our
model and construct the KSBEs. In Sec.~III, we lay out our main numerical
results. The effects of electric fields, together with the carrier density,
temperature and intervalley scattering (note that in this paper, 
with the term ``intervalley  scattering'', we always indicate the $\Gamma$-$L$
intervalley scattering unless otherwise specified)  on the SRT are discussed. We
summarize in Sec.~IV. 

\section{Model and KSBEs}
We start our investigation in $n$-type bulk GaAs where the four $L$
valleys locate at $L$ points [${\bf  K}^0_{L_i}=\frac{\pi}{a_0}(1,\pm 1,\pm 1)$ with
$a_0$ denoting the lattice constant and $i=1$-$4$] and lie energetically $E_{\Gamma
  L}=0.296$~eV above the $\Gamma$ valley.\cite{Vurga} 
It is noted that the four $L$ valleys can be arbitrarily chosen from the eight
$L$ points limited by the condition that there is no center inversion symmetry
between any two of them. In the spherically symmetric 
approximation, the electron effective masses of the $\Gamma$ and $L$ valleys are
represented as $m_\Gamma=0.067m_0$ and $m_L=0.23m_0$,\cite{Lei1,Chand} respectively, with
$m_0$ representing the free electron mass. We do not consider valleys 
of higher energy, e.g., the next-nearest valleys $X$ which are $E_{\rm LX}=0.166$~eV above
the $L$ valleys, since even for the highest electric field ($E=8$~kV/cm) employed in
this work, the fractions of electrons in these valleys are negligible.\cite{Kratzer}

The KSBEs derived from the nonequilibrium Green function method
reads\cite{Wu1,Wu2,Wu3,Zhang1} 
\begin{equation}
\left. \left. \left. \partial_t \rho_{\lambda{\bf k}_\lambda}=
\partial_t\rho_{\lambda{\bf k}_\lambda}\right|_{\rm coh}+\partial_t\rho_{\lambda{\bf k}_\lambda}\right|_{\rm drift}
+\partial_t\rho_{\lambda{\bf k}_\lambda}\right|_{{\rm scat}}, \label{KSBE1}
\end{equation}
in which $\rho_{\lambda{\bf k}_\lambda}$ is the density matrix of electrons with momentum
${\bf k}_\lambda$ in $\lambda$ valley. Note that ${\bf k}_\lambda$ is defined in reference to
the valley center in each valley. The diagonal term $\rho_{\lambda{\bf
    k}_\lambda,\sigma\sigma}\equiv f_{\lambda{\bf k}_\lambda,\sigma}$ ($\sigma=\pm
1/2$) describes the
distribution of each spin band and the off-diagonal term $\rho_{\lambda{\bf
    k}_{\lambda},\frac{1}{2}-\frac{1}{2}}=\rho_{\lambda{\bf
    k}_{\lambda},-\frac{1}{2}\frac{1}{2}}^\ast$ is the correlation between the
two spin bands. The coherent term is given by
\begin{equation}
\left. \partial_t\rho_{\lambda{\bf k}_\lambda}\right|_{\rm coh}=-i\Big[{\bf \Omega}_{\lambda}({\bf
  k}_\lambda)\cdot{\bm \sigma}+\Sigma^{\lambda}_{\rm HF}({\bf
  k}_\lambda),\rho_{\lambda{\bf k}_\lambda}\Big], \label{Coh}
\end{equation}
where $[\ ,\ ]$ is the commutator and ${\bf \Omega}_{\lambda}({\bf
  k}_\lambda)$ represents the Dresselhaus SOC in $\lambda$ valley.\cite{Dresselhaus} 
By setting the [100] and [010] directions as $x$- and $y$-axises, respectively,
${\bf \Omega}_{\Gamma}({\bf k}_\Gamma)$ takes the form\cite{Dresselhaus,Ivchenko2} 
\begin{eqnarray}
\nonumber
  {\bf \Omega}_{\Gamma}({\bf k}_\Gamma) &=& \gamma_D [k_{\Gamma x}(k_{\Gamma
    y}^2-k_{\Gamma z}^2), k_{\Gamma y}(k_{\Gamma z}^2-k_{\Gamma x}^2), \\ &&k_{\Gamma
    z}(k_{\Gamma x}^2-k_{\Gamma y}^2)].\label{DresselG}
\end{eqnarray}
By further denoting $\hat{\bf n}_{1/3}=(1,1,\pm 1)/\sqrt{3}$ and  $\hat{\bf n}_{2/4}=-(1,\pm 1,
1)/\sqrt{3}$, which are the unit vectors of the longitudinal principle axis
of the $L_i$ valleys, we have for the $L$ valleys\cite{Dresselhaus,Ivchenko2,Zhang1}
\begin{equation}
  {\bf \Omega}_{L_i}({\bf k}_{L_i}) = \beta_D (k_{L_{i x}}, k_{L_i y}, k_{L_i
    z})\times \hat{{\bf n}}_i.\label{DresselL}
\end{equation}
$\Sigma^\lambda_{\rm HF}({\bf k}_\lambda)=-\sum_{{\bf
    k}_\lambda^{\prime}}V_{\lambda \lambda,{\bf k}_\lambda-{\bf
k}^{\prime}_\lambda}\rho_{\lambda{\bf k}_\lambda^{\prime}}$ in Eq.~(\ref{Coh})
is the Coulomb Hatree-Fock  term with $V_{\lambda \lambda,{\bf k}_\lambda-{\bf 
k}^{\prime}_\lambda}$ representing the intravalley Coulomb scattering matrix
element. The drift term takes the form 
$\left. \partial_{t}\rho_{\lambda{\bf k}_\lambda}\right|_{\rm drift} = e {\bf
  E}\cdot\mbox{\boldmath$\nabla$\unboldmath}_{{\bf k}_\lambda} \rho_{\lambda{\bf  
    k}_\lambda}$ ($e>0$).  $\left. \partial_{t}\rho_{\lambda{\bf k}_\lambda}\right|_{\rm
  scat}$ stands for the scattering terms, which include intra- and
inter-valley parts with the explicit expressions given in Appendix~A. 

\section{Numerical Results}
In this section, we present our results obtained by numerically solving the
KSBEs following the scheme laid out in Refs.~\onlinecite{Jiang1} and
\onlinecite{Zhang1}. All parameters used in our computation are listed in
Table~I. 

The initial spin polarized state of the system is prepared in the similar way as in
Ref.~\onlinecite{Zhang1}, starting from the unpolarized equilibrium state. That
state is first driven to the drifted steady state under the electric field. The
main difference lies that after driving the system to the 
steady state, we turn on a circularly polarized laser pulse at $t_1=6$~ps to
excite spin polarized electrons with a degree of injected spin polarization $P_{\rm
  inject}=50\%$ into the $\Gamma$ valley:\cite{Meier,Korn} $\delta
f_{\Gamma{\bf k}_{\Gamma},\sigma}$=$\alpha_\sigma\exp[-(\varepsilon^\Gamma_{{\bf 
  k}_{\Gamma}}-\varepsilon_{\rm
pump})^{2}/2\delta_{\varepsilon}^{2}][1-f_{\Gamma{\bf
  k}_{\Gamma},\sigma}(t_1)]$. Here $\alpha_\sigma=n_{{\rm pump},\sigma}/\{\sum_{{\bf k}_{\Gamma}}
\exp[-(\varepsilon^\Gamma_{{\bf k}_{\Gamma}}-\varepsilon_{\rm pump})^{2}/2
\delta_{\varepsilon}^{2}][1-f_{\Gamma{\bf k}_{\Gamma},\sigma}(t_1)]\}$. $\varepsilon_{\rm pump}$ is
the energy of pulse center in reference to the band minimum and
$\delta_{\varepsilon}$=$\hslash/\delta_{\tau}$ 
with $\delta_{\tau}$ denoting the pulse width. $n_{\rm pump,\sigma}$ is the
density of electrons with spin $\sigma$
after excitation with $n_{\rm pump,\frac{1}{2}}=3n_{\rm pump,-\frac{1}{2}}$. In
this work,  we employ $\varepsilon_{\rm
  pump}=4$~meV for the case of 
resonant excitation, $\delta_{\tau}=0.01$~ps and $n_{\rm pump}=n_{\rm
  pump,\frac{1}{2}}+n_{\rm pump,-\frac{1}{2}}=0.02\times n_e$. Here $n_e$ stands
for the unpolarized electron density before 
pumping which is equal to the doping density, so the total spin polarization after
the pump pulse is $P_0\approx 1\%$. It is noted that due to the 
strong electron-electron Coulomb scattering, the drifted steady-state distribution is
established within $0.1$~ps after the pump pulse. Therefore the pulse
characters, i.e., $\varepsilon_{\rm pump}$ and $\delta_{\tau}$, have little 
influence on the SRT. It is further noted that due to the small initial spin
polarization, the exact value of the polarization has marginal effect on the
SRT.\cite{Jiang1} 

\begin{figure}[htb]
  \begin{center}
    \includegraphics[width=7cm]{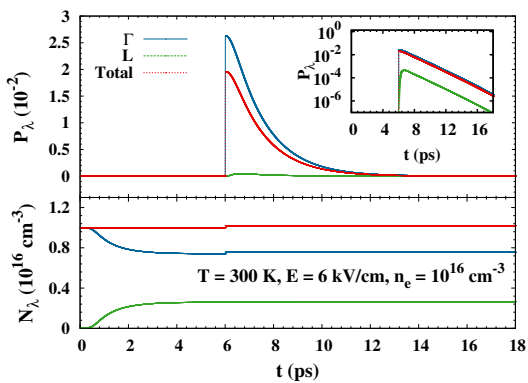}
  \end{center}
  \caption{(Color online) Typical evolution of the electron spin polarization
    and population in $\lambda$ valley in the condition with
    $n_{e}=10^{16}$~cm$^{-3}$, $E=6$~kV/cm and $T=300$~K. } 
  \label{fig:time}
\end{figure}

In Fig.~\ref{fig:time}, we plot the typical time evolution of the spin
polarization and also the electron population in $\lambda$ valley in the
condition with electric field $E=6$~kV/cm, electron density
$n_{e}=10^{16}$~cm$^{-3}$ and temperature $T=300$~K. Clear 
transfer of electrons from the $\Gamma$ valley to the $L$ valleys is observed.
In order to quantitatively understand the influence of the  $L$
valleys, we calculate the steady-state 
drift velocities of $\lambda$-valley electrons, the mobilities, the electron
populations and the hot-electron temperatures in the $\Gamma$ and $L$ valleys as
function of electric field. The 
explicit results are plotted in Appendix~B with underlying physics
analyzed. From Fig.~\ref{fig:All-E}(a) of Appendix~B, one finds that our model is validated
against the experimental data.\cite{Kratzer}

In the following, we focus on how the SRT is influenced by the electric field,
carrier density and lattice temperature. The effect of 
the $L$ valleys is also studied. The SRT $\tau$ is obtained by fitting
temporal evolution of spin polarization $P_{\lambda}$ with an exponential decay
(see the inset of Fig.~\ref{fig:time}). Throughout this work, the electric field is
applied in the $x$-direction and its strength is limited up to 8~kV/cm where the
$X$ valleys can be neglected.\cite{Kratzer}

\begin{table}[htbp]
\caption{Material parameters used in the calculation (from
  Refs.~\onlinecite{Zhang1} and \onlinecite{Vurga} unless otherwise specified). }
\begin{tabular}{llllllllll}\hline\hline
 $E^{\Gamma}_g$ (eV)  &\ \ \ &  1.519  &\ \ \ \ \ \ &  $\Omega_{\rm \Gamma\Gamma}$ (meV) &\ \ \ &  35.4  &&  \\ 
 $E^{L}_g$ (eV)      &&  1.815  &&   $\Omega_{\rm L_iL_i}$ (meV) &&  34.3  &&      \\
 $E^{X}_g$ (eV)      &&  1.981  &&       $\Omega_{\rm \Gamma L}$ (meV) &&  20.8  &&      \\
$m^*_\Gamma/m_0$         &&  0.067  &&    $\Omega_{\rm L_i L_j}$ (meV) &&  29.0  &&      \\
$m^*_L/m_0$          &&  0.23  &&     $D_{\rm L_i L_i}$ ($10^9$~eV/cm) &&  0.3  &&      \\
$\kappa_0$         &&  12.9  &&    $D_{\rm \Gamma L}$ ($10^9$~eV/cm) &&  1.1  &&      \\
$\kappa_{\infty}$  &&  10.8  &&       $D_{\rm L_i L_j}$ ($10^9$~eV/cm) &&  1.0  &&      \\
$\gamma_{\rm D}$ (eV$\cdot$\AA$^3$) &&  23.9$^a$ &&    $d$  ($10^3$~kg/m$^{3}$)   && 5.36 && \\
$\beta_{\rm D}$ (eV$\cdot$\AA) &&  0.26 &&   &&&&\\
\hline\hline
\end{tabular}\\
\hspace{-6.3cm}$^a$ Ref.~\onlinecite{Fu}. 
\end{table}

\subsection{Electric field  dependence and effect of $L$ valleys}
\begin{figure}[htb]
  \begin{center}
    \includegraphics[width=7cm]{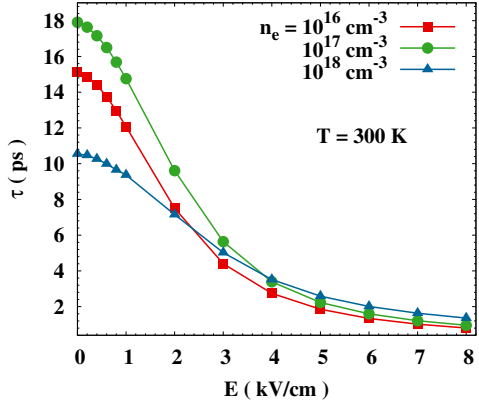}
  \end{center}
  \caption{(Color online) SRT $\tau$ {\em vs.} electric field
    $E$ at temperature $T=300$~K for three electron densities: $n_e=10^{16}$,
    $10^{17}$ and $10^{18}$~cm$^{-3}$. }  
  \label{fig:t-E}
\end{figure}

We first study the electric field dependence of the SRT at $T=300$~K. In
Fig.~\ref{fig:t-E}, we plot the SRTs against electric field for 
electron densities $n_e=10^{16}$, $10^{17}$ and $10^{18}$~cm$^{-3}$,
corresponding to the nondegenerate, intermediate and 
degenerate regimes in the field-free condition, respectively. It is seen
that the SRT is effectively modulated by the electric field. For all three
cases, the SRTs decrease monotonically with the electric field and
reach down to the ones with one order of magnitude smaller than the
corresponding field-free values at $E=8$~kV/cm. This is different 
from the previous work in $n$-type GaAs quantum wells 
where a nonmonotonic electric field 
dependence is observed,\cite{Weng1,Weng3,Zhang1} but consistent with the very
recent work in bulk GaAs by Jiang and 
Wu where the electric field is applied up to 2~kV/cm.\cite{Jiang1} In the regime
with $E < 2$~kV/cm, according to Fig.~\ref{fig:All-E}(c) in Appendix~B, the $L$
valleys are still irrelevant. Therefore the underlying physics of the monotonic
decrease of the SRT is the same as that analyzed in Ref.~\onlinecite{Jiang1}:  
Due to the cubic form of the Dresselhauss
SOC in bulk, the enhancement of inhomogeneous broadening from the drift
effect and the hot-electron effect [see Fig.~\ref{fig:All-E}(d) in Appendix~B where
the two--hot-electron-temperature behavior of $\Gamma$-valley electrons is discussed] is
more pronounced than that in quantum 
wells with small well width where the SOC is in the linear
form.\cite{Weng1,Weng3,Zhang1} It hence overtakes the 
enhancement of momentum scattering and leads to the monotonic field dependence
of the SRT. However, when the electric field is increased over 
2~kV/cm, a visible amount of electrons are drifted into the $L$ valleys [see
Fig.~\ref{fig:All-E}(c) in Appendix~B] and hence the $L$ valleys are expected to
play a role in the total spin relaxation.  

\begin{figure}[htb]
  \begin{minipage}[]{10cm}
    \hspace{-0.5 cm}\parbox[t]{8cm}{
      \includegraphics[width=6cm]{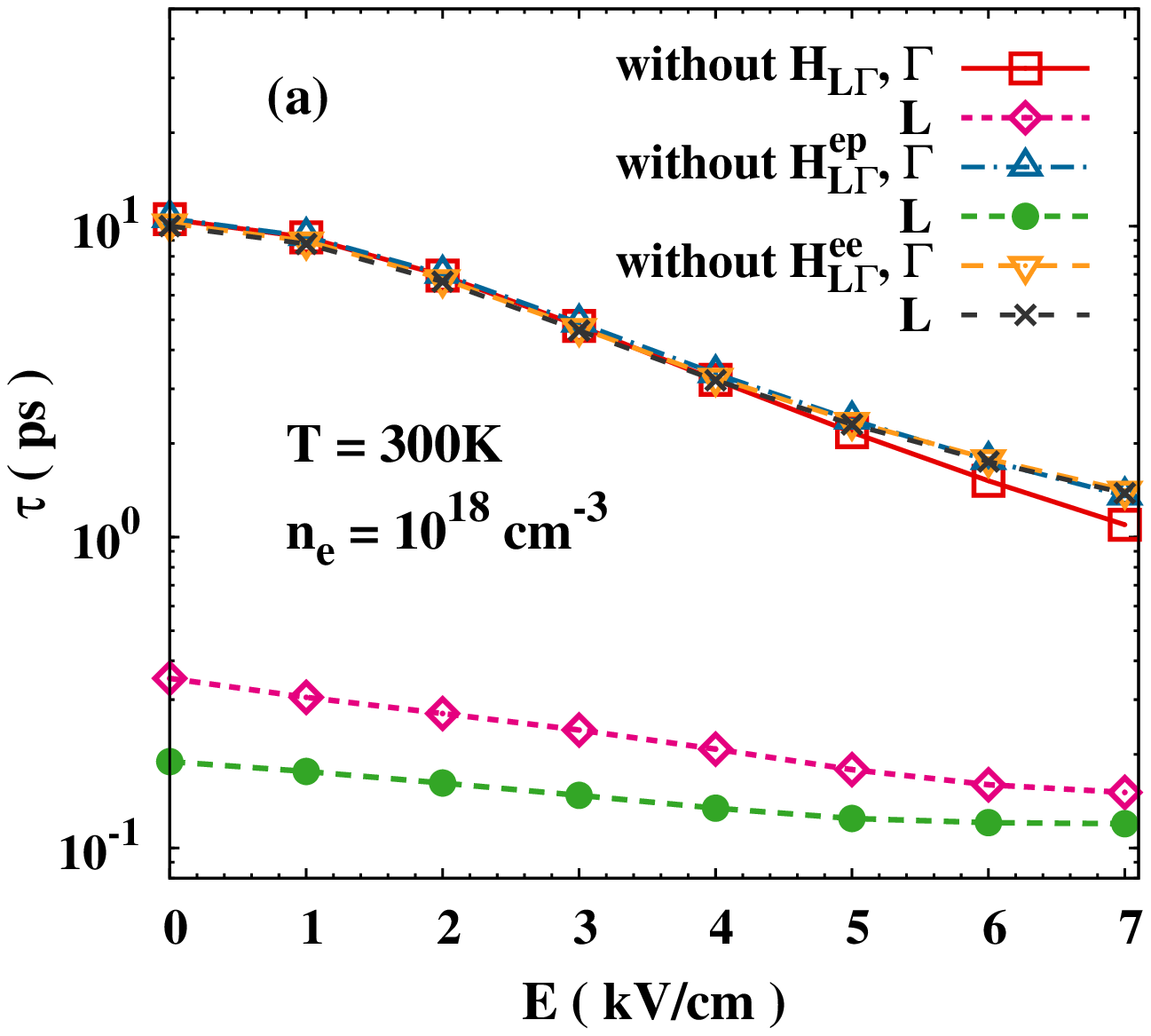}}
  \end{minipage}
  \begin{minipage}[]{10cm}
    \hspace{-1.5 cm}\parbox[t]{5cm}{
      \includegraphics[width=4.5cm,height=4.3 cm]{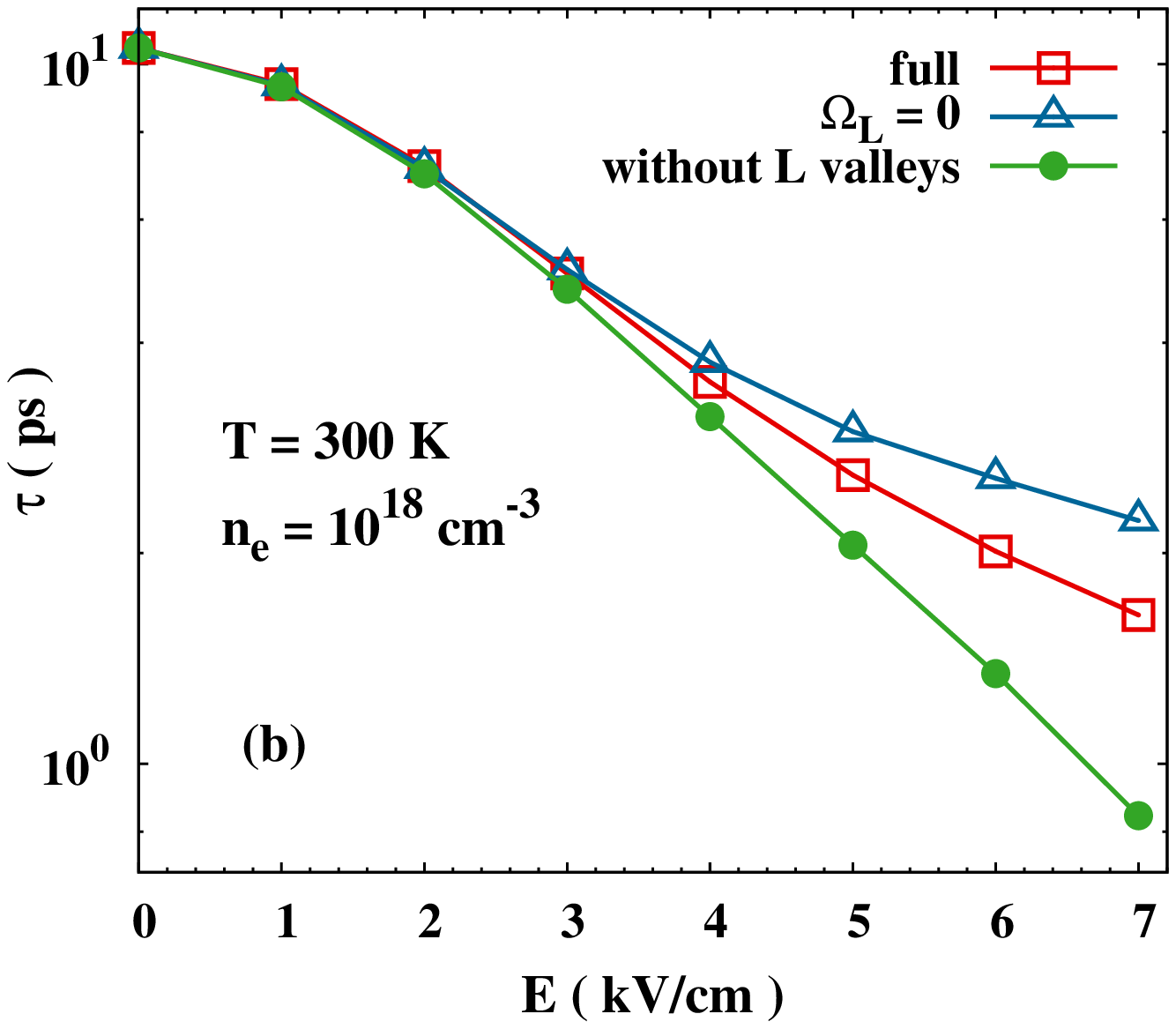}}
    \hspace{-0.7 cm}\parbox[t]{5cm}{
      \includegraphics[width=4.8cm,height=4.3 cm]{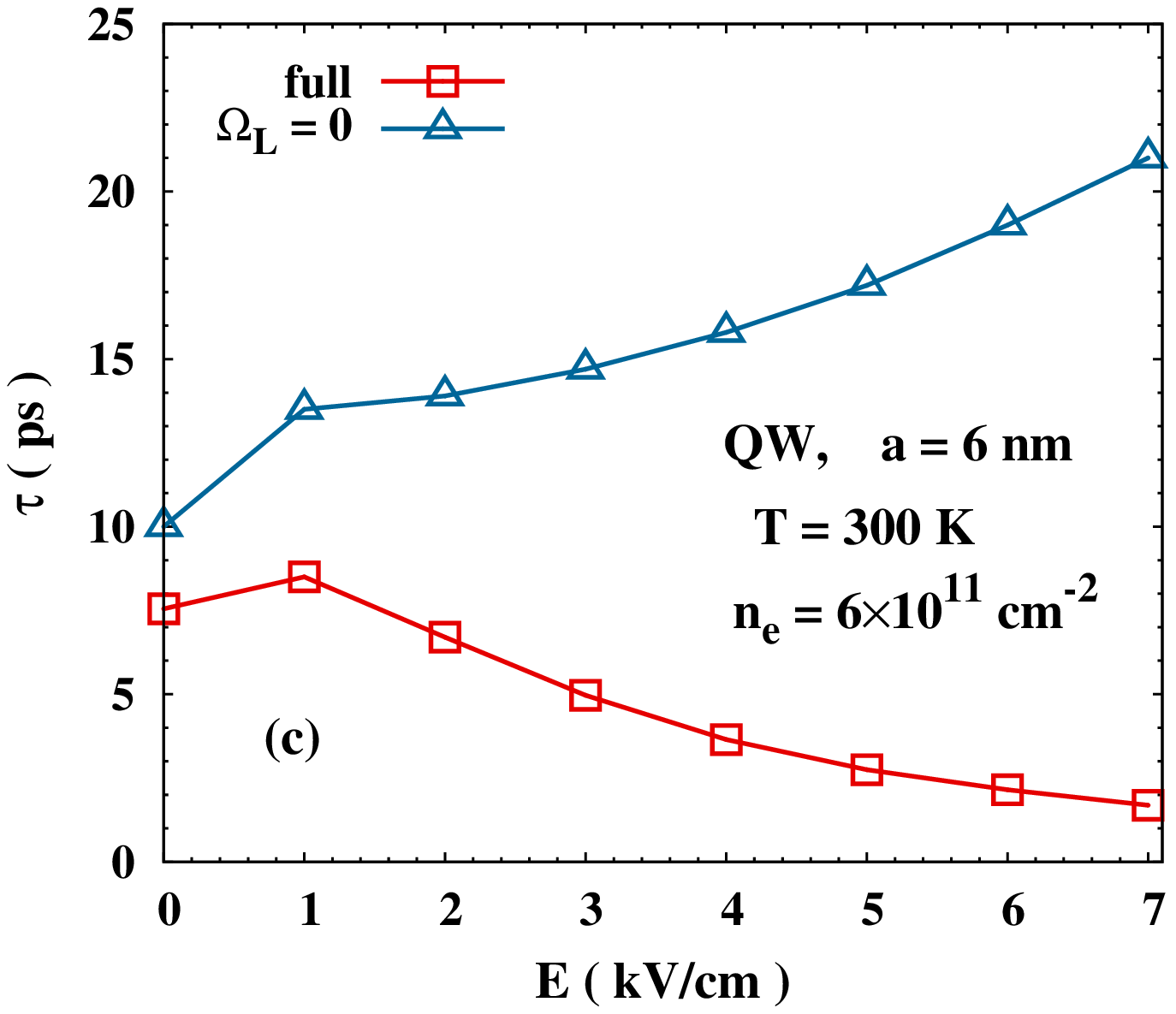}}
  \end{minipage}
  \begin{minipage}[]{10cm}
    \hspace{-1.5 cm}\parbox[t]{5cm}{
      \includegraphics[width=4.5cm,height=4.3 cm]{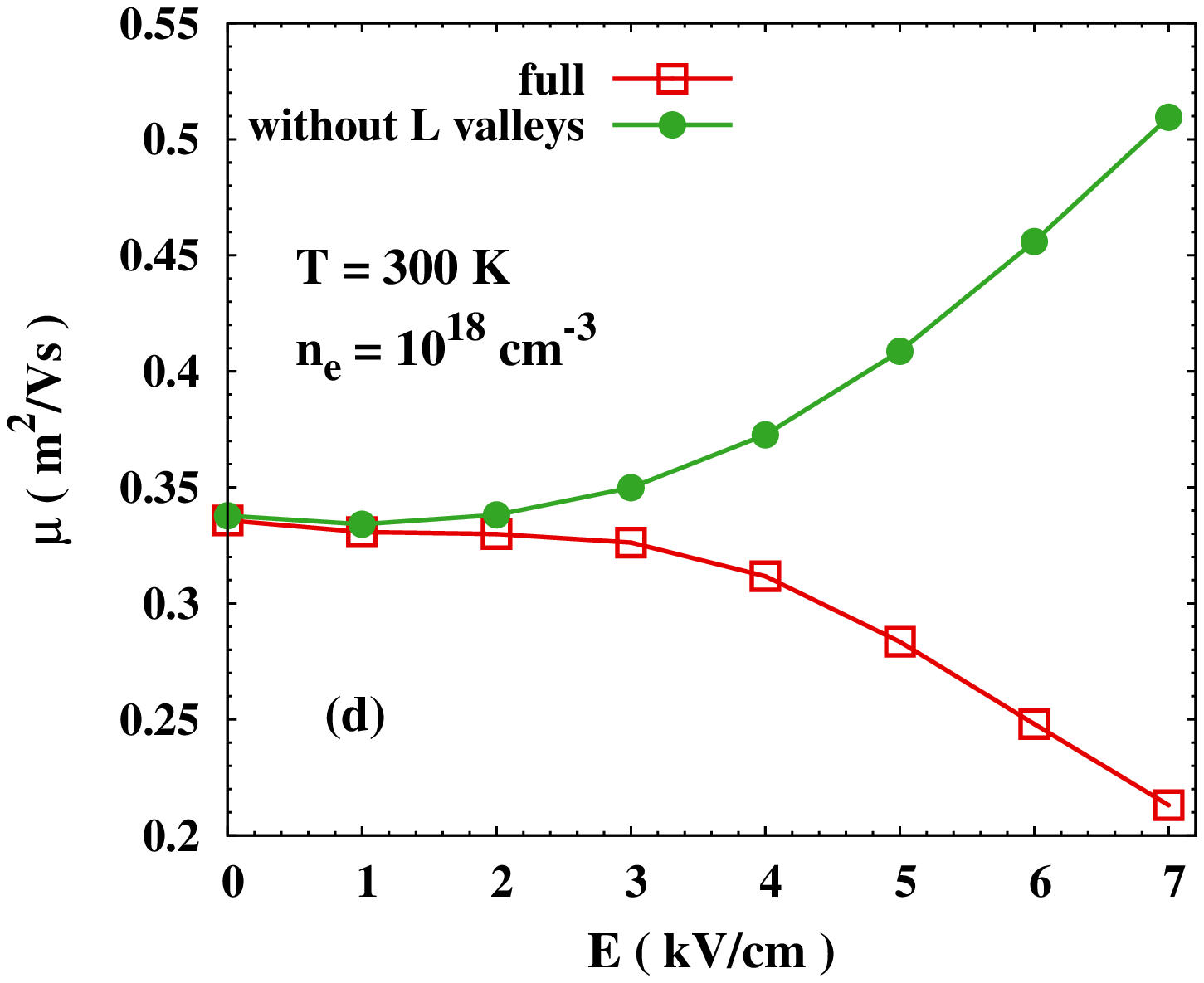}}
    \hspace{-0.7 cm}\parbox[t]{5cm}{
      \includegraphics[width=4.5cm,height=4.3 cm]{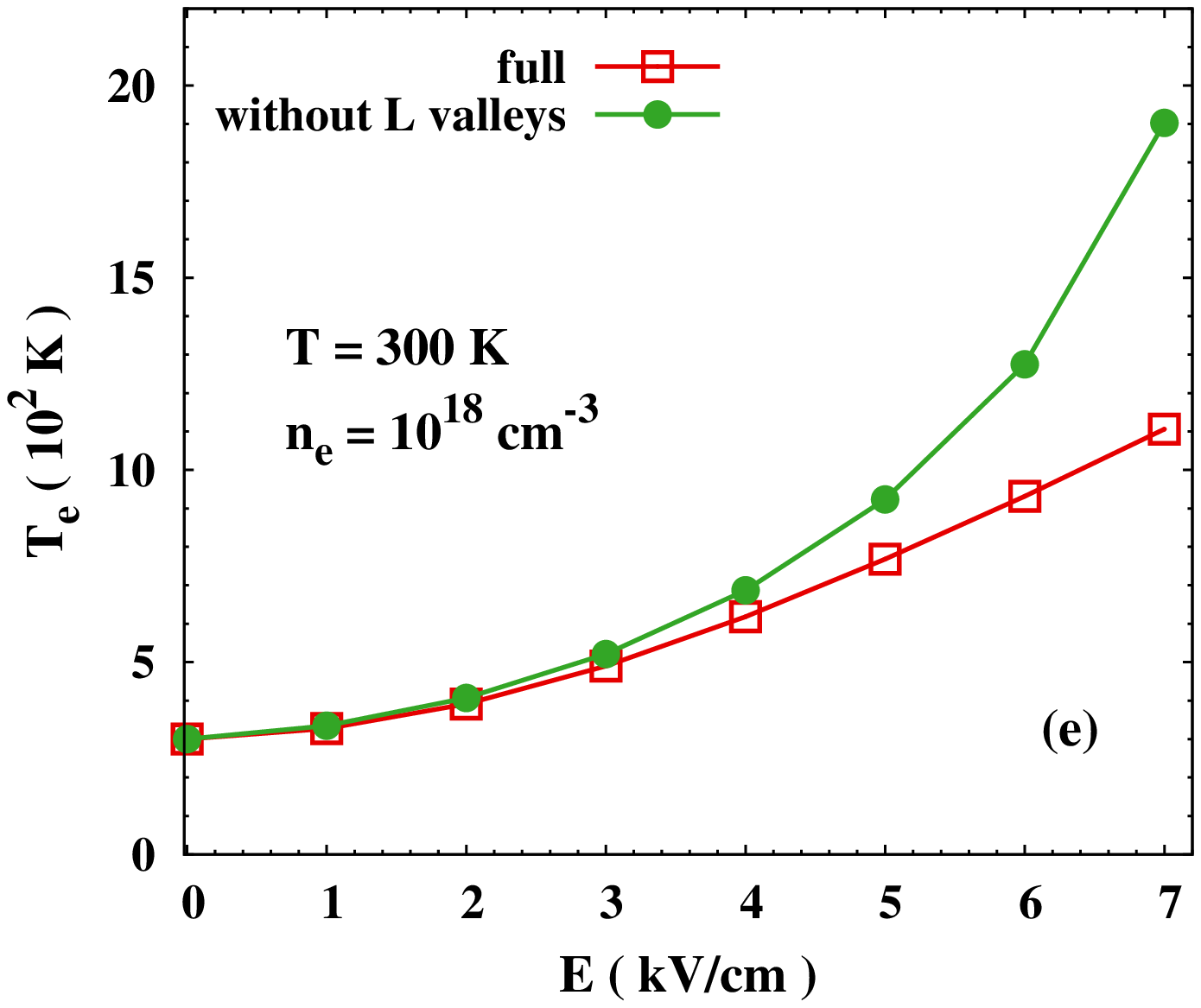}}
  \end{minipage}
  \caption{(Color online) For (a), (b), (d) and (e): in bulk with electron density
    $n_e=10^{18}$~cm$^{-3}$:  (a) SRTs $\tau$ 
    of the $\Gamma$ and  $L$ valleys as function of electric field $E$ by only removing
    the intervalley scattering $H_{L\Gamma}$ (curves with $\square$ and 
    $\diamond$), the intervalley electron-phonon scattering $H_{L\Gamma}^{\rm ep}$ (curves
    with $\triangle$ and $\bullet$) or the intervalley electron-electron Coulomb 
    scattering $H_{L\Gamma}^{\rm ee}$ (curves  with $\triangledown$ and $\times$) . (b) SRT $\tau$ {\em
      vs.} electric field $E$ in the 
    genuine condition [Case (i), curve with $\square$], by setting the SOC in
  the $L$ valleys to zero [Case (iii), curve with $\triangle$] and without 
    the $L$ valleys [Case (ii), curve with $\bullet$]. (d) and (e) Mobility $\mu$
    and  hot-electron temperature 
    $T_\Gamma$ {\em vs.} electric field with [Case (i), curve with $\square$] and 
    without [Case (ii), curve with $\bullet$] the $L$ valleys. For (c) in quantum
    well (QW) with well width $a=6$~nm and $n_e=6\times 10^{11}$~cm$^{-2}$, SRT
    $\tau$ {\em vs.} electric field $E$ in the genuine condition [Case (i),
    curve with $\square$],  and by setting the SOC in the $L$ valleys to zero [Case (iii),
    curve with $\triangle$]. $T=300$~K in all the situations.} 
  \label{fig:L1}
\end{figure}

An important feature of the multivalley spin relaxation 
can be noticed in the inset of
Fig.~\ref{fig:time} that the evolutions of spin
polarizations in the $\Gamma$ and $L$ valleys share the same damping rate. 
This can be understood in respect to the rapid exchange of electrons between the
$\Gamma$ and $L$ valleys resulting from the strong intervalley scattering. As
shown in Fig.~\ref{fig:L1}(a),  by removing the intervalley scattering
$H_{\Gamma L}$, one finds  that the spin polarizations in separate valleys now
evolve independently and the ``intrinsic'' SRT in the $L$ valleys
(curves with $\diamond$) becomes one order of magnitude smaller than that
of the $\Gamma$ valley (curve with $\square$). This demonstrates the crucial   
role of the intervalley scattering in obtaining the identical damping of spin
polarizations in the $\Gamma$ and $L$ valleys. 
To be more specific, we  also calculate the SRT in each valley by
removing only the intervalley electron-phonon or  electron-electron Coulomb 
scattering. It is seen from Fig.~\ref{fig:L1}(a) that the SRTs in the $\Gamma$ and
$L$ valleys without the intervalley electron-electron Coulomb scattering coincide,
whereas those without the intervalley electron-phonon scattering are far away
from each other. This demonstrates that the identical damping of spin
polarizations in the $\Gamma$ and $L$ valleys comes from the intervalley
electron-phonon scattering. It is noted that the fast spin
relaxation in the $L$ valleys hints that they may serve as a ``drain'' of the
total spin polarization, just as the case in quantum wells.\cite{Zhang1}  

To further elucidate the role of the $L$ valleys in the bulk system, we
calculate the SRTs by artificially 
removing the $L$ valleys [labeled as Case (ii)] and by setting the SOC
in the $L$ valleys to zero [labeled as Case (iii)] in the computation. The calculated
SRTs are compared to those obtained in the genuine condition
[labeled as Case (i)]. 
In Fig.~\ref{fig:L1}(b), we plot the SRTs as function of electric field for Case
(i)-(iii) with $n_e=10^{18}$~cm$^{-3}$ 
and $T=300$~K. One observes that for $E<2$~kV/cm, the effect of the $L$ valleys
is marginal due to little electrons drifted into the $L$
valleys under weak electric field. When the electric field further
increases, the $L$ valleys start to play 
an important role and obvious distinctions are seen for the three cases. 
We first compare the genuine SRT with the one without the $L$ valleys.
As shown in Fig.~\ref{fig:L1}(b), by removing the $L$ valleys in the
calculation, surprisingly the SRT (curve with $\bullet$) becomes smaller instead
of larger than its genuine value (curve with $\square$). For
$E=7$~kV/cm, the SRT without the $L$ valleys reaches down to {\em half} of the
genuine value. In order to understand this behavior, we calculate 
the mobility and the hot-electron temperature $T_\Gamma$, which serve as
scales of the drift and hot-electron effect, respectively. From
Fig.~\ref{fig:L1}(d) and (e), pronounced 
distinctions are seen for $E>4$~kV/cm. In the case without the $L$ 
valleys, the mobility increases rather than decreases with the electric field
and reaches more than twice of its genuine value at $E=7$~kV/cm. Meanwhile, the
hot-electron temperature $T_\Gamma$ reaches almost twice of its genuine
value. Consequently, both the drift and hot-electron effects are markedly
enhanced without the $L$ valleys, which in turn leads to a drastic increase of
the inhomogeneous broadening due to the cubic form the
$\Gamma$-valley Dresselhaus SOC in bulk.
 It overtakes the effect
of the absence of the $L$ valleys and leads to the smaller SRT compared to the 
case with the $L$ valleys. This tells us that the $L$ valleys serve as a ``momentum
damping area'' where electrons from the $\Gamma$ valley are blocked from  reaching
higher momentum states, thanks to the large effective mass in
the $L$ valleys.

We then focus on the curves corresponding to Case (i) (curve with $\square$) and
(iii) (curve with $\triangle$). The SRT calculated by setting the SOC in the $L$
valleys to zero becomes a little larger and reaches less than $3/2$ of the corresponding
genuine value at $E=7$~kV/cm, indicating that the spin relaxation in the
$L$ valleys only slightly modulates the total spin relaxation. This is very 
different from the quantum well system. As a comparison, in
Fig.~\ref{fig:L1}(c) we plot the SRTs against electric field in a quantum
well in Case (i) and (iii).\cite{Zhang1,note2-QW} The well width is chosen as $6$~nm
and the two-dimensional electron density is $n_e=6\times 10^{11}$~cm$^{-2}$.  It is
seen that the behavior of field dependence of the SRT is totally 
changed in the absence of the SOC in the $L$ valleys. Instead of decreasing rapidly after a
small increase, the SRT with ${\bf \Omega}_L=0$ 
increases monotonically and reaches
more than ten times of the corresponding genuine value at $E=7$~kV/cm. 

The underlying physics is understood as follows: in bulk, due to the cubic form of the
Dresselhauss SOC in the $\Gamma$ valley, the field-induced drift effect and the
hot-electron effect lead to a stronger enhancement of the inhomogeneous broadening
compared to the increase of momentum scattering. Moreover, a simple
calculation shows that although in average, the inhomogeneous broadening in
the $\Gamma$ valley is much smaller than that in the $L$ valleys, in the
energy range of the $\Gamma$ valley overlapping with the $L$ valleys (which is
roughly where the exchange of electrons between the $\Gamma$ and $L$ valleys
happens), the effective magnetic field from the Dresselhauss SOC is already 
comparable with that in the $L$ valleys. However, in quantum
wells with a small well width, the enhancement of scattering is more effective
than that of the inhomogeneous broadening thanks to the linear form of the
Dresselhaus SOC.\cite{Zhang1} Besides, the effective magnetic field in the
energy range of the $\Gamma$ valley overlapping with the $L$ valleys is still much
smaller than that  in the $L$ valleys. In addition, it is further noted that
electrons are more liable to be drifted into the $L$ valleys in quantum
wells. In Fig.~\ref{fig:L1}(c), $n_L/n_e$ reaches 45\% at $E=7$~kV/cm compared
to 25\% in bulk, which further enhances  the effect of the $L$ valleys in
quantum wells. Nevertheless, in Fig.~\ref{fig:L1}(c), the electric field
corresponding to  $n_L/n_e=0.25$ is $E=5$~kV/cm, at which the SRT without the 
SOC in the $L$ valleys is still about five times larger
  than the corresponding genuine 
value. All these lead to the pronounced difference between two- and three-dimensional
systems. It also tells us that the $L$ valleys no longer serve as ``spin drain'' in
bulk, differing from the case of quantum wells.\cite{Zhang1} 

\subsection{Density dependence}
\begin{figure}[htb]
  \begin{center}
    \includegraphics[width=7cm]{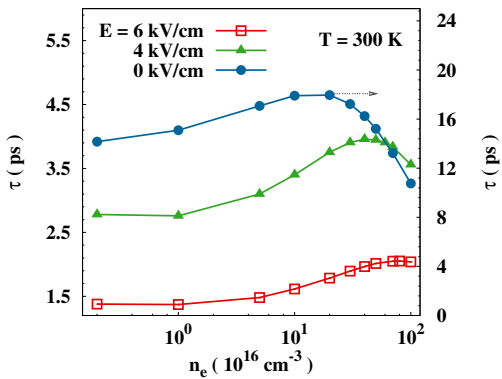}
  \end{center}
  \caption{(Color online) SRT $\tau$ {\em vs.} electron
    density at temperature $T=300$~K for fixed electric field: $E=0$, 4 and
    6~kV/cm. Note that the scale of the case $E=0$~kV/cm is on the right hand side of the frame.} 
  \label{fig:t-N}
\end{figure}

Another interesting feature seen in Fig.~\ref{fig:t-E} is that the $\tau$-$E$ curve
corresponding to the electron density $n_e=10^{18}$~cm$^{-3}$ intersects with
the other two, indicating that the density dependence of the SRT changes with the
variation of electric  field. In order to elucidate this behavior, we calculate
the SRT by varying the electron density $n_e$ from $2\times 10^{15}$ to 
$10^{18}$~cm$^{-3}$ with electric field $E=0$, 4 and
6~kV/cm. Note that in bulk the impurity density $n_i$ is always taken as $n_e$
in this paper. From Fig.~\ref{fig:t-N}, one 
observes that for $E=0$~kV/cm, there is a peak at around $n_e=2.0\times
10^{17}$~cm$^{-3}$. This peak in the density dependence of the SRT has been
theoretically predicted\cite{Jiang1} and experimentally
confirmed\cite{Cundiff,Shen1,Bob,MengY} very recently, and is attributed to the
crossover from the nondegenerate to degenerate limit when the
electron-impurity and electron-electron Coulomb scatterings are the dominant
scattering processes.\cite{Jiang1} The crossover between degenerate and
nondegenerate limits can be estimated by the Fermi
temperature $T_F$ [$=(3\pi^2n_e)^{2/3}/(2m)$], with the peak determined by $T_F\sim
\frac{1}{2}T$-$T$.\cite{Jiang1,Cundiff,Shen1,Bob,Mad,Jiang2,BYsun} Here the Fermi
temperature corresponding to the electron density at the peak is $T_F\approx
207$~K, comparable to the  lattice temperature $T=300$~K ($T_F/T\sim 2/3$),
in line with the previous works.  

By applying an electric field $E=4$~kV/cm, 
one observes from Fig.~\ref{fig:t-N} that the peak moves to around
$n_e=5\times 10^{17}$~cm$^{-3}$, and when the electric field is further increased to
$E=6$~kV/cm, the peak is shifted to even higher electron density (at around
$n_e=9\times 10^{17}$~cm$^{-3}$ hence is not very obvious in the 
figure). We point out that this is due to the increase of hot-electron
temperature induced by the strengthened electric field. As pointed out by 
Shen,\cite{Shen1} because of the
laser-induced hot-electron effect, the peak of density dependence of the SRT
appears at where the hot-electron temperature $T_e$, rather than the lattice
temperature $T$, approximately equals the
Fermi temperature $T_F$. The underlying 
physics is similar here with the electric-field--induced hot-electron
effect. However, as shown in Fig.~\ref{fig:All-E} in Appendix~B, the $L$ valleys
start to play a role with high electric field and moreover, the 
two--hot-electron-temperature behavior of the $\Gamma$-valley electrons is seen.  
This makes the situation more complex. Nevertheless, by comparing the curves of 
``intrinsic'' SRTs of the $\Gamma$ and $L$ valleys in Fig.~\ref{fig:L1}(a) with the 
corresponding genuine one in Fig.~\ref{fig:L1}(b), one notices that the 
curve of the ``intrinsic'' SRT in the $\Gamma$ valley resembles the genuine one
while that of the  ``intrinsic'' SRT in the $L$ valleys is far away it. This
indicates that the the multivalley spin relaxation in 
the presence of high electric field is mainly determined by the $\Gamma$
valley. Furthermore, since most electrons stay in the $\Gamma$ valley for
electric field up to 6~kV/cm [see Fig.~\ref{fig:All-E}(c) in Appendix~B] and
these electrons mostly distribute in the lower-energy regime compared to the
high-energy regime overlapping the $L$ valleys, the behavior of the SRT is
mainly determined by this part of electrons.   
In the approximation that all electrons are in the $\Gamma$ valley, for
$E=4$~kV/cm, we have $T_F\approx 381$~K according to the electron density at
the peak, comparable to the corresponding hot-electron temperature $T_e\approx
659$~K ($T_F/T_e\sim 3/5$). Meanwhile for $E=6$~kV/cm, we have $T_F\approx
563$~K compared to $T_e\approx 981$~K at the peak ($T_F/T_e\sim 3/5$).  We note
that this effect of electric field on the density dependence of the SRT can be
observed within current technology of optical orientation. 

\subsection{Temperature dependence}
\begin{figure}[htb]
  \begin{minipage}[]{10cm}
    \hspace{-1.5 cm}\parbox[t]{8cm}{
      \includegraphics[width=5cm]{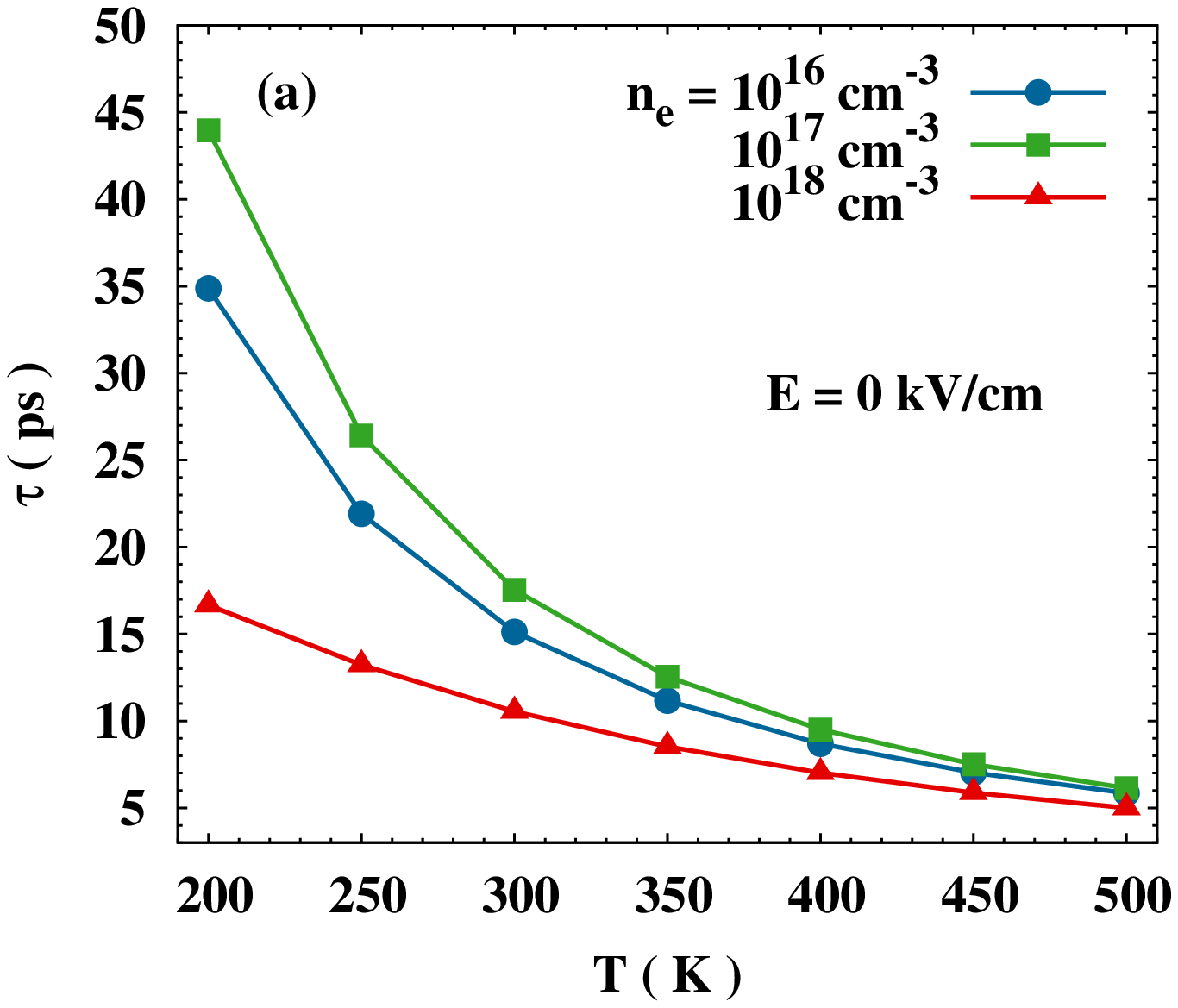}}
  \end{minipage}
  \begin{minipage}[]{10cm}
    \hspace{-1.5 cm}\parbox[t]{5cm}{
      \includegraphics[width=4.6cm,height=4.3 cm]{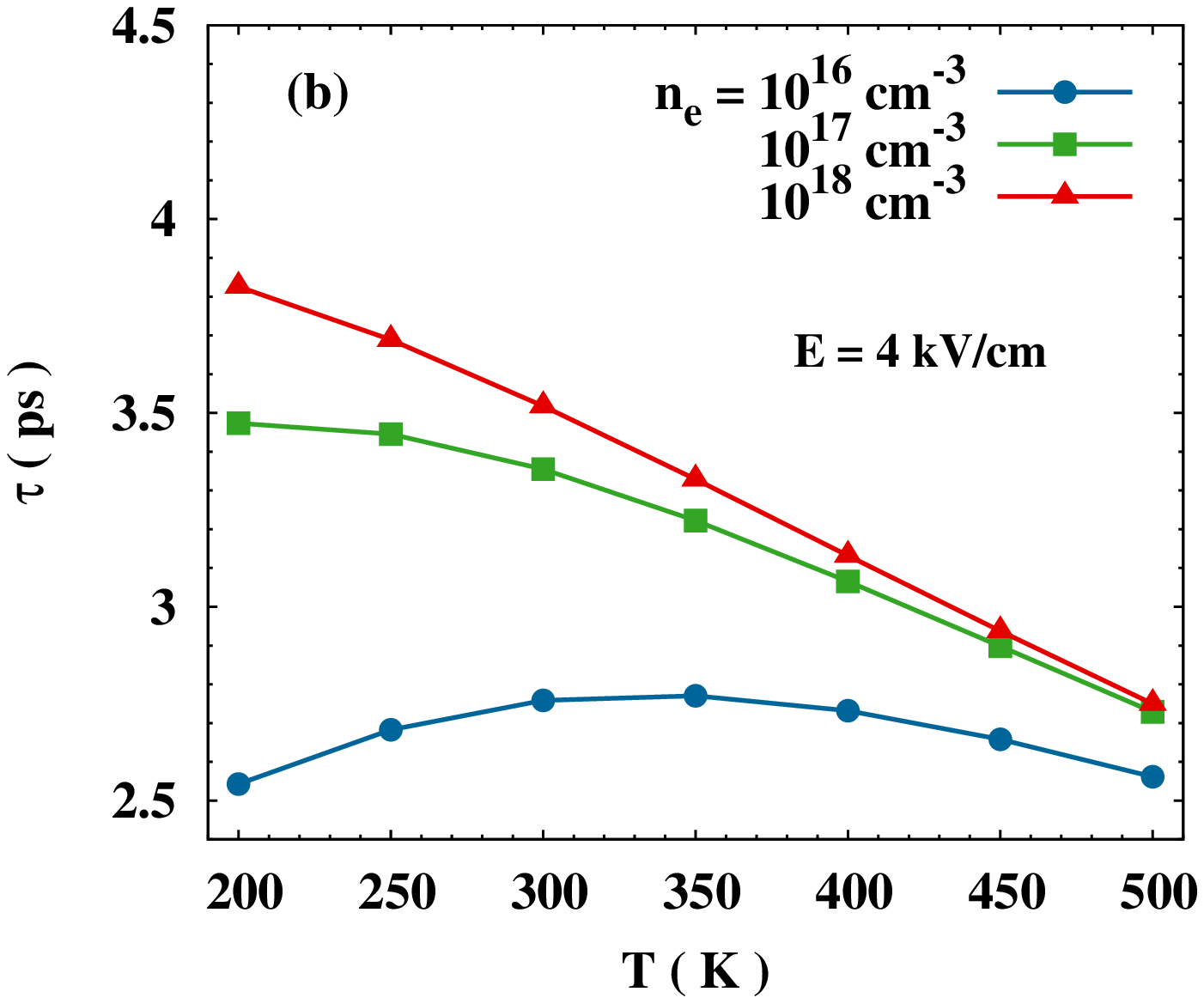}}
    \hspace{-0.7 cm}\parbox[t]{5cm}{
      \includegraphics[width=4.6cm,height=4.3 cm]{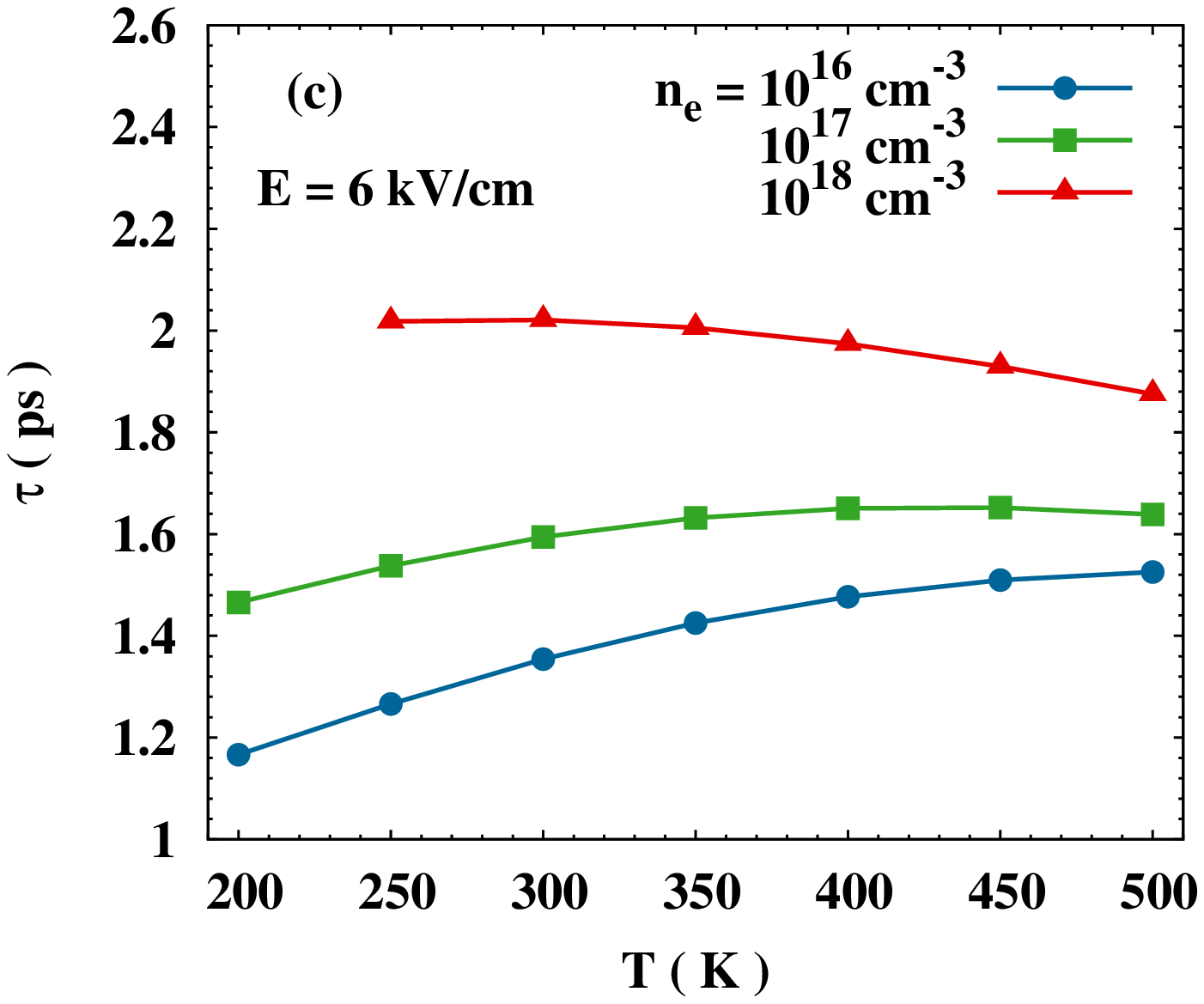}}
  \end{minipage}
  \begin{minipage}[]{10cm}
    \hspace{-1.5 cm}\parbox[t]{5cm}{
      \includegraphics[width=4.5cm,height=4.2 cm]{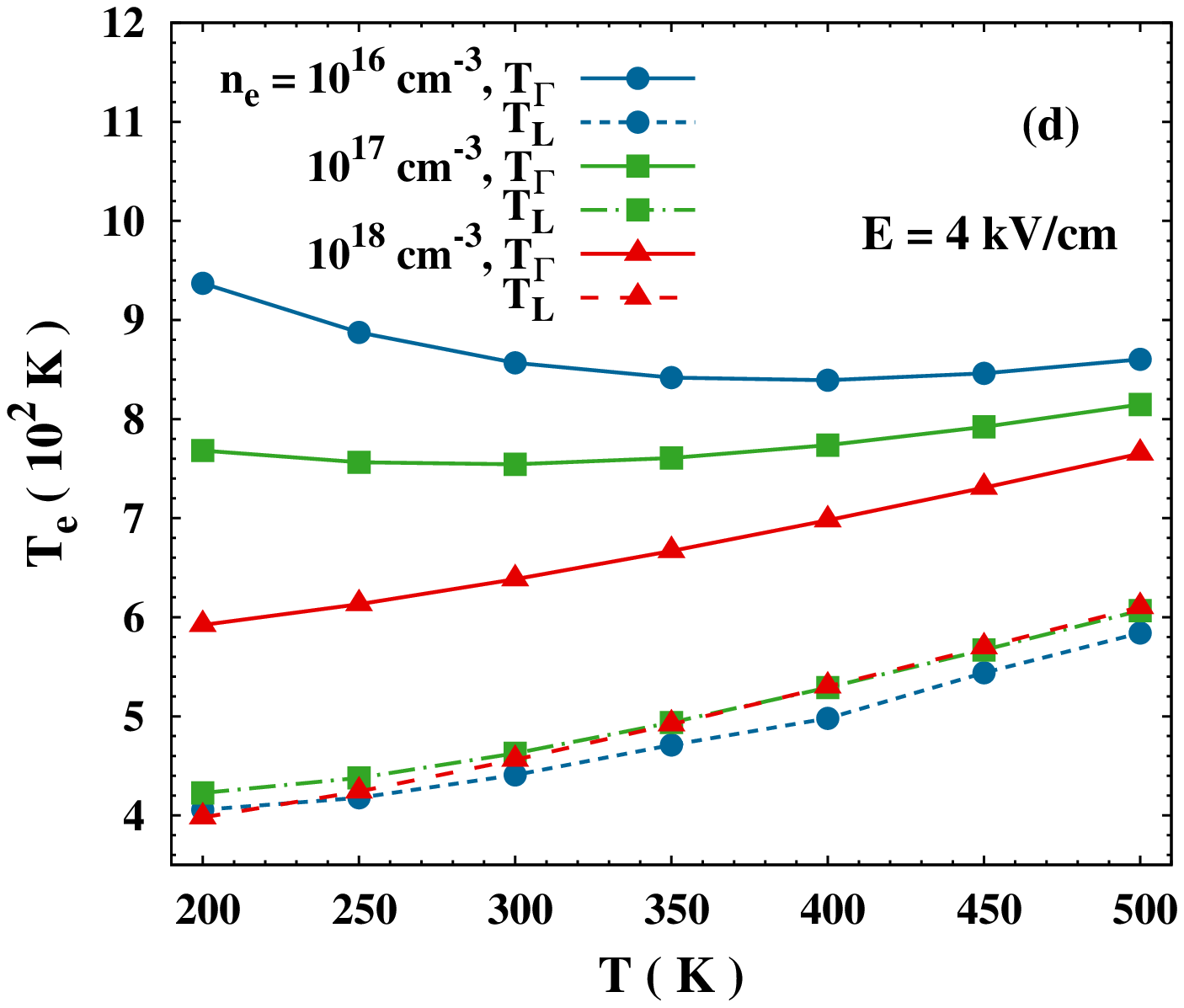}}
    \hspace{-0.7 cm}\parbox[t]{5cm}{
      \includegraphics[width=4.5cm,height=4.2 cm]{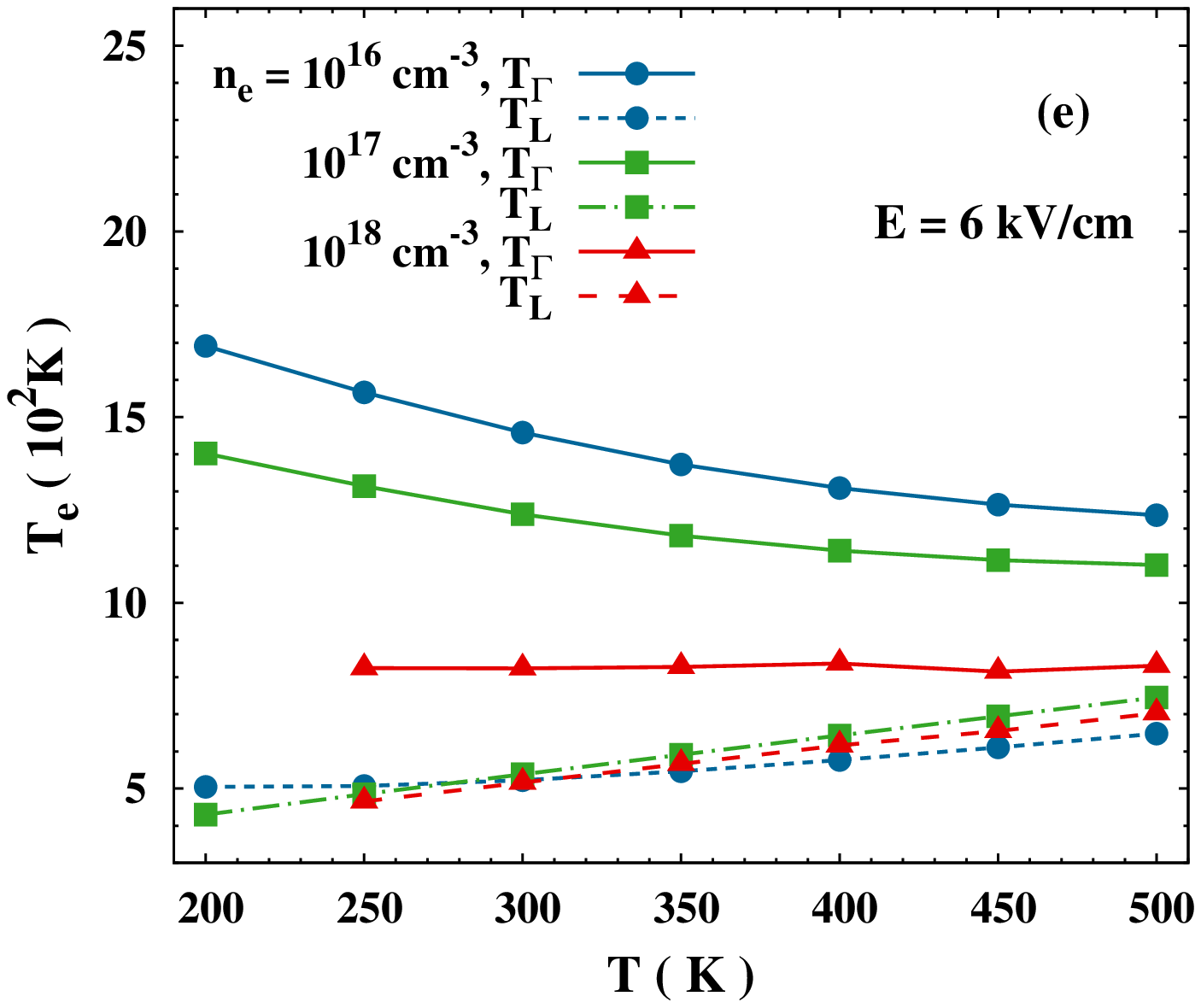}}
  \end{minipage}
  \caption{(Color online) SRT $\tau$ {\em vs.} lattice
    temperature $T$ with (a) $E=0$~kV/cm; (b) $E=4$~kV/cm and (c)
    $E=6$~kV/cm. Hot-electron temperatures $T_\Gamma$ and $T_L$ {\em vs.} lattice
    temperature $T$ with (d) $E=4$~kV/cm and (e) $E=6$~kV/cm. The electron
    densities are $n_e=10^{16}$, $10^{17}$ and $10^{18}$~cm$^{-3}$ in these figures. }  
  \label{fig:t-T}
\end{figure}

We now turn to investigate the temperature dependence of the SRT.  In
Fig.~\ref{fig:t-T}, the SRTs obtained are plotted as function of lattice temperature
with electric field $E=0$, 4 and 6~kV/cm and for 
three electron densities as in Sec.~IIIA. In the field-free case,
it has been formerly shown both experimentally\cite{Kikkawa,Murdin,Oertel,Beschoten} and
 theoretically\cite{Jiang1,Zhou1} that in $n$-type samples 
with low mobility, the SRT decreases monotonically with temperature. From
Fig.~\ref{fig:t-T}(a), it is seen that our result coincides 
with the previous ones. However, very different behaviors are seen under high
electric fields. Apart from the overall suppressed values of the SRT compared to
the field-free condition, it is
found that for electron density $n_e=10^{18}$~cm$^{-3}$, the decreasing  
rate becomes smaller for higher electric fields. Whereas for
$n_e=10^{17}$~cm$^{-3}$, the SRT turns to increase with increasing temperature under 
electric field $E=6$~kV/cm. The most interesting phenomenon is seen for the case
with $n_e=10^{16}$~cm$^{-3}$. From Fig.~\ref{fig:t-T}(a)-(c), one
observes that the SRT decreases monotonically 
with increasing lattice temperature $T$ in the absence of the electric field but
increases monotonically with it when electric field $E=6$~kV/cm is applied. In
between, for the case with $E=4$~kV/cm, the SRT first increases and then
decreases with increasing temperature $T$, with a peak at around $T=350$~K.   
This peak in the temperature dependence of the SRT is very different from the one
theoretically predicted by Zhou {\em et al}.\cite{Zhou1} and experimentally
realized by Leyland {\em et al}.,\cite{Leyland} Ruan {\em et al}.\cite{Ruan} and
Han {\em et al}.\cite{MengY} for high-mobility samples in the field-free
condition. There, the peak is solely caused by the electron-electron Coulomb 
scattering\cite{Wu1,Weng2,Brand,Ivchenko1,Bronold,Zhou1,Leyland,Ruan,MengY} and
appears in the crossover between the degenerate and nondegenerate limits where
the Fermi temperature $T_F$ is comparable to the lattice
temperature.\cite{Zhou1,Leyland,Ruan,MengY} However, in $n$-type bulk 
materials, due to the strong electron-impurity scattering (hence low mobility),
the Coulomb scattering is always less important and no peak is expected 
in the temperature dependence of the SRT.\cite{Jiang1,Murdin,Oertel}
Moreover, the Fermi temperature corresponding to $n_e=10^{16}$~cm$^{-3}$ is
$T_F\approx 28$~K, which is
far below the lattice temperature, let alone the hot-electron temperature
under the high electric field. This further demonstrates the essential
difference of the peak observed here.

This complex behavior of the SRT can be understood
from the different lattice-temperature dependences of the hot-electron
temperature at the different electron densities and electric
fields. Corresponding to Fig.~\ref{fig:t-T}(b) and (c), in Fig.~\ref{fig:t-T}(d) and
(e) we plot the hot-electron temperatures against the lattice temperature for
$E=4$ and 6~kV/cm, respectively. Note that the same color and type of point are
used for the corresponding electron density.  By comparing the
$T_\Gamma$-$T$ curves in Fig.~\ref{fig:t-T}(b) with the $\tau$-$T$ ones in
Fig.~\ref{fig:t-T}(d) and those in Fig.~\ref{fig:t-T}(c) with 
Fig.~\ref{fig:t-T}(e), direct correspondence of the temperature dependence of the
SRT to that of the hot-electron temperature $T_\Gamma$ is observed. The order
of curves from top to bottom reverses in the corresponding two figures as higher
hot-electron temperature indicates larger inhomogeneous broadening, and hence
smaller SRT. Meanwhile, the decrease (increase) of the hot-electron
temperature gives an increase (decrease) of the SRT in the lattice
temperature dependence. Specifically, the
monotonic decrease (increase) of the SRT with $n_e=10^{18}$~cm$^{-3}$
($n_e=10^{16}$~cm$^{-3}$) in Fig.~\ref{fig:t-T}(b) [Fig.~\ref{fig:t-T}(c)]  
corresponds to the monotonic increase (decrease) of the hot-electron
temperature in Fig.~\ref{fig:t-T}(d) [Fig.~\ref{fig:t-T}(e)]. Especially
corresponding to the peak in Fig.~\ref{fig:t-T}(b) for $n_e=10^{16}$~cm$^{-3}$,
there is a valley in the lattice-temperature dependence of the 
hot-electron temperature in Fig.~\ref{fig:t-T}(d). 
For the electric field-free case, the electron temperature equals the lattice
temperature, so the behavior of the SRT in Fig.~\ref{fig:t-T}(a) is also
in the same trend as the high field cases. It is noted that in the discussion
above, we focus on $T_\Gamma$ instead of $T_L$. This is because, as discussed
in the previous subsection, the total spin relaxation is 
mainly determined by the $\Gamma$ valley while most $\Gamma$-valley electrons
stay in the lower-energy regime. Meanwhile, according to the the discussions
in Sec.~IIIA, in the variation of the hot-electron 
temperature, the variation of inhomogeneous broadening is more profound than
that of the momentum relaxation, thanks to the cubic ${\bf
  k}$ dependence of the $\Gamma$-valley Dresselhauss
SOC.\cite{Dresselhaus,Ivchenko2} Therefore the variation of the inhomogeneous
broadening mainly determines the variation of the SRT. 

We point out that the different behaviors of hot-electron temperature
with the lattice temperature at different electron densities and electric
fields are originated from the temperature dependence of the 
energy-gain and loss rates of the electron system.\cite{Conwell,Seeger,Lei3,Lei2}
With the increase of lattice temperature, the mobility is reduced resulting from
the enhancement of scattering, so does the the energy-gain rate; meanwhile the
energy-loss rate decreases due to increase of 
phonon number.\cite{Conwell,Seeger,Lei3,Lei2} These two effects compete with each other and
lead to the complex behavior of the hot-electron temperature in different
conditions. A more detailed discussion is  given in Appendix~C. 

\section{Summary}
In summary, we have investigated the multivalley spin relaxation in $n$-type bulk
GaAs in the presence of high electric field by applying 
the fully microscopic KSBE approach. The $\Gamma$ and $L$ valleys, 
which are relevant in determining the properties of spin dynamics for the high electric
field applied in this work, are taken into account. The effect of the $L$ valleys on spin
relaxation is discussed and is shown to be very different from the quantum
well system. We find that apart from the effect of directly manipulating the SRT, the high
electric field can also effectively modulate the density and temperature
dependences of the SRT.

First, the SRT is found to decrease monotonically with the electric field for electron
densities from the nondegenerate to degenerate limit. This
monotonic field dependence of the SRT is very different from the previous works in
$n$-type GaAs quantum wells\cite{Weng1,Weng3,Zhang1} and is assigned to the
pronounced enhancement of inhomogeneous broadening from the field-induced drift
and hot-electron effects thanks to the cubic form of the $\Gamma$-valley
Dresselhauss SOC in bulk. 
We show that, in despite of the very different strength of the SOC in different valleys, the
evolutions of spin polarizations in the $\Gamma$ and $L$ valleys share the {\em same}
damping rate. This is demonstrated to come from the strong
intervalley electron-phonon scattering and indicates the feasibility of
exploring the properties of the hot electrons in the $L$ valleys through the phenomena detected
in the $\Gamma$ valley. Moreover, differing from the role of a ``spin drain'' of
the total spin polarization in the two-dimensional  system, we find that in bulk
the $L$ valleys serve as a ``momentum damping 
area'' where electrons are blocked from drifting to higher momentum states due to
the large effective mass. This tends to suppress the inhomogeneous broadening 
and in turn leads to a {\rm longer} rather than shorter SRT compared to
the case without the $L$ valleys.

As for the density dependence
of the SRT, the formerly predicted\cite{Jiang1} and 
experimentally observed\cite{Cundiff,Bob} density peak in the field-free
condition is recovered and is found to be shifted to higher density regime with higher
electric field. We attribute this to the electric-field--induced hot-electron
effect.

We also investigate the temperature dependence of the SRT in conditions with different electron
densities and electric fields.  The monotonic decrease of the SRT with increasing
lattice temperature in the field-free condition coincides with the
previous works. Nevertheless, the SRT is found to decrease more slowly with higher
electric field, and even turn to increase monotonically with increasing lattice
temperature in the condition with low electron density and high electric field. 
More interestingly, a peak is predicted in the temperature dependence with low
electron density and moderate electric field, which is vastly different from the
formerly discussed one in high mobility samples in the field-free condition. We
point out that this peculiar behavior of the SRT originates from the 
temperature dependence of the energy-gain and loss rates of the electron system. 
 
Finally, we remark on the feasibility of the spin Gunn effect in $n$-type bulk
GaAs. For the preferred electron density in Ref.~\onlinecite{Flatte} 
($n_e=10^{18}$~cm$^{-3}$), we note that the SRT is suppressed down to the value
shorter than what required for the spontaneous spin amplification to appear
under the electric fields where the charge Gunn effect appears. This fast
damping of spin polarization overtakes the  
spontaneously generation process and makes the spin Gunn effect hardly to be
realized in $n$-type bulk GaAs system.

\begin{acknowledgments}
This work was supported by
the National Natural Science Foundation of China under Grant No. 10725417 and
the National Basic Research 
Program of China under Grant No. 2012CB922002. Discussions with X. Marie is
acknowledged. We also
would like to thank P. Zhang for calculating the SRTs in quantum wells in
Fig.~\ref{fig:L1}(c). One of the authors (H.T.) 
would like to thank K. Shen for valuable discussions.
\end{acknowledgments}

\begin{appendix}
\section{Scattering Terms in KSBEs}
The scattering term $\left. \partial_{t}\rho_{\lambda{\bf k}_\lambda}\right|_{{\rm
    scat}}$ includes the contributions from the
electron-impurity scattering $\left. \partial_{t}\rho_{\lambda{\bf k}_\lambda}\right|_{{\rm
    ei}}$, the electron-phonon scattering
$\left. \partial_{t}\rho_{\lambda{\bf k}_\lambda}\right|_{{\rm ep}} $ and the
electron-electron Coulomb scattering $\left. \partial_{t}\rho_{\lambda{\bf k}_\lambda}\right|_{{\rm ee}}$
\begin{equation}
\left.\partial_{t}\rho_{\lambda{\bf k}_\lambda}\right|_{{\rm scat}}=
\left.\partial_{t}\rho_{\lambda{\bf k}_\lambda}\right|_{{\rm ei}} +
\left.\partial_{t}\rho_{\lambda{\bf k}_\lambda}\right|_{{\rm ep}} +
\left.\partial_{t}\rho_{\lambda{\bf k}_\lambda}\right|_{{\rm ee}},  
\label{scat_all}
\end{equation}
where
\begin{eqnarray}
\nonumber
\hspace{-0.5cm} &&\left.\partial_{t}\rho_{\lambda{\bf k}_\lambda}\right|_{{\rm ei}}\\ 
\nonumber
&& = - \pi n_i Z_i^2\sum_{{\bf k}_\lambda^{\prime}}V_{{\bf k}_\lambda-{\bf
    k}_\lambda^{\prime}}^2 \delta(\varepsilon^{\lambda}_{{\bf 
    k}_\lambda^{\prime}}-\varepsilon^\lambda_{{\bf k}_\lambda})  \\ 
&&\hspace{0.3cm} \mbox{}\times\Big(\rho^{>}_{\lambda{\bf
    k}_\lambda^{\prime}}\rho^{<}_{\lambda{\bf k}_\lambda} - \rho^{<}_{\lambda{\bf
    k}_\lambda^{\prime}}\rho^{>}_{\lambda{\bf k}_\lambda} \Big)  + {\rm h.c.},
\label{scat_ei}\\
\nonumber
&&\left.\partial_{t}\rho_{\lambda{\bf k}_\lambda}\right|_{{\rm ep}}\\ 
\nonumber 
&&= - \pi\sum_{\lambda^\prime,{\bf
    k}^\prime_{\lambda^{\prime}},\pm} |M_{\lambda\lambda^\prime,{\bf k}_\lambda-{\bf
    k}^\prime_{\lambda^{\prime}}}|^2\delta(\pm\Omega_{\lambda\lambda^{\prime}}+\varepsilon^{\lambda^\prime}_{{\bf  
    k}^{\prime}_{\lambda^{\prime}}}-\varepsilon^\lambda_{{\bf k}_\lambda}) \\ 
&&\hspace{0.3cm}\mbox{} \times
  \Big(N_{\lambda\lambda^{\prime}}^{\pm} \rho^{>}_{\lambda^\prime{\bf k}^{\prime}_{\lambda^{\prime}}}
    \rho^{<}_{\lambda{\bf k}_\lambda} -
    N_{\lambda\lambda^\prime}^{\mp}\rho^{<}_{\lambda^\prime{\bf
        k}^{\prime}_{\lambda^{\prime}}} 
    \rho^{>}_{\lambda{\bf k}_\lambda} \Big) + {\rm h.c.} , 
\label{scat_ep}\\
\nonumber
&&\left.\partial_{t}\rho_{\lambda{\bf k}_\lambda}\right|_{{\rm ee}} \\
\nonumber
&&= - \pi \sum_{\lambda^\prime,{\bf k}_\lambda^{\prime},{\bf
    k}_{\lambda^{\prime}}^{\prime\prime}} \delta(\varepsilon^{\lambda}_{{\bf
    k}_\lambda^{\prime}}-\varepsilon^\lambda_{{\bf
    k}_\lambda}+\varepsilon^{\lambda^\prime}_{{\bf
    k}_{\lambda^{\prime}}^{\prime\prime}}-\varepsilon^{\lambda^\prime}_{{\bf 
    k}_{\lambda^{\prime}}^{\prime\prime}-{\bf k}_\lambda+{\bf
    k}_\lambda^{\prime}}) \\ 
\nonumber
&& \hspace{0.3cm}\mbox{}\times V_{{\bf k}_\lambda-{\bf
    k}_\lambda^{\prime}}^2 \Big[ \rho^{>}_{\lambda{\bf
    k}_\lambda^{\prime}}\rho^{<}_{\lambda{\bf k}_\lambda} {\rm
  Tr}\left(\rho^{<}_{\lambda^\prime({\bf
      k}_{\lambda^{\prime}}^{\prime\prime}-{\bf 
      k}_\lambda+{\bf k}_\lambda^{\prime})}\rho^{>}_{\lambda^\prime{\bf
      k}_{\lambda^{\prime}}^{\prime\prime}}\right) 
\nonumber\\
&&\hspace{0.3cm} \mbox{} - \rho^{<}_{\lambda{\bf
    k}_\lambda^{\prime}}\rho^{>}_{\lambda{\bf k}_\lambda}{\rm
  Tr}\left(\rho^{>}_{\lambda^\prime({\bf
      k}_{\lambda^{\prime}}^{\prime\prime}-{\bf 
      k}_\lambda+{\bf k}_\lambda^{\prime})}\rho^{<}_{\lambda^\prime{\bf
      k}_{\lambda^{\prime}}^{\prime\prime}}\right) \Big]+ {\rm h.c.}. \ \ \ 
\label{scat_ee}
\end{eqnarray}
In these equations, $\rho_{\bf k}^{<}=\rho_{\bf k}$ and $\rho_{\bf
  k}^{>}=1-\rho_{\bf
  k}$. $n_i$ is the impurity density which equals the electron density in this
paper and $Z_i=1$ is the charge number of the
impurity. $\varepsilon^\Gamma_{{\bf k}_\Gamma}=k_\Gamma^2/(2m_\Gamma^*)$ and
$\varepsilon^{L_i}_{{\bf k}_{L_i}}=k_{L_i}^2/(2m_{L}^*)+\varepsilon_{\Gamma L}$ with
$\varepsilon_{\Gamma L}$ denoting the energy difference between the $\Gamma$ and $L$
points. $N_{\lambda\lambda^{\prime}}=[e^{(\Omega_{\lambda\lambda^{\prime}}/T)}-1]^{-1}$   
is the Bose distribution of phonons with frequency
$\Omega_{\lambda\lambda^\prime}$. $N^<_{\lambda\lambda^\prime}=N_{\lambda\lambda^\prime}$ and 
$N^>_{\lambda\lambda^\prime}=1+N_{\lambda\lambda^\prime}$.
$M_{\lambda\lambda^{\prime},{\bf q}}$ is the matrix element of the 
electron-phonon scattering with ${\bf q}$ standing for the phonon wave
vector. Here we take into account the intra- and intervalley
electron--longitudinal-optical (LO) phonon scattering in and between the $\Gamma$ and
$L$ valleys, respectively.\cite{Zhang1,Lei1,Birman,Herbert,Fawcett,Mick} For the
intravalley electron-phonon scattering, we have $M_{\Gamma\Gamma,{\bf
    q}}^{2}=\frac{e^{2}\Omega_{\Gamma\Gamma}(\kappa_{\infty}^{-1}-\kappa_{0}^{-1})}{2\epsilon_{0}q^{2}}$
and $M_{L_i L_i,{\bf q}}^{2}=\frac{D_{L_iL_i}^{2}}{2d\Omega_{L_i L_i}}$. Also we
have $M_{\Gamma L_{i},{\bf q}}^{2}=M_{L_{i}\Gamma,{\bf
    q}}^{2}=\frac{D_{\Gamma L}^{2}}{2d\Omega_{\Gamma L}}$ for
the $\Gamma$-$L$ intervalley electron-phonon scattering and $M_{L_{i}L_{j},{\bf q}}^{2}=\frac{
  D_{L_{i}L_{j}}^{2}}{2d\Omega_{L_{i}L_{j}}}$ for $L$-$L$
intervalley electron-phonon scattering. $V_q$ is the screened Coulomb potential
under the random phase approximation,\cite{Mahan}
\begin{equation}
  V_{\bf q} = \frac{V^{(0)}_{\bf q}}{1-V^{(0)}_{\bf q}P^{(1)}({\bf
  q})}, 
\end{equation}
where
\begin{equation}
  P^{(1)}({\bf q}) = \sum_{\lambda,{\bf k}_\lambda,\sigma} \frac{f_{\lambda({\bf k}_\lambda+{\bf
  q}),\sigma}-f_{\lambda{\bf k}_\lambda,\sigma}}{\varepsilon^\lambda_{{\bf k}_\lambda+{\bf
  q}}-\varepsilon^\lambda_{{\bf k}_\lambda}},
\end{equation}
with $V^{(0)}_{\bf q} = e^2/(\epsilon_0\kappa_0q^2)$ denoting the bare Coulomb
potential and $f_{\lambda{\bf k},\sigma}$ being the electron distribution of spin-$\sigma$
band. It is noted that in Eq.~(\ref{scat_ee}), we include both the intra- and
  intervalley electron-electron Coulomb scattering.
 All parameters appear in these equations are listed in Table.~I.

\begin{figure}[htb]
  \begin{minipage}[]{10cm}
    \hspace{-1.5 cm}\parbox[t]{5cm}{
      \includegraphics[width=4.5cm,height=4.5 cm]{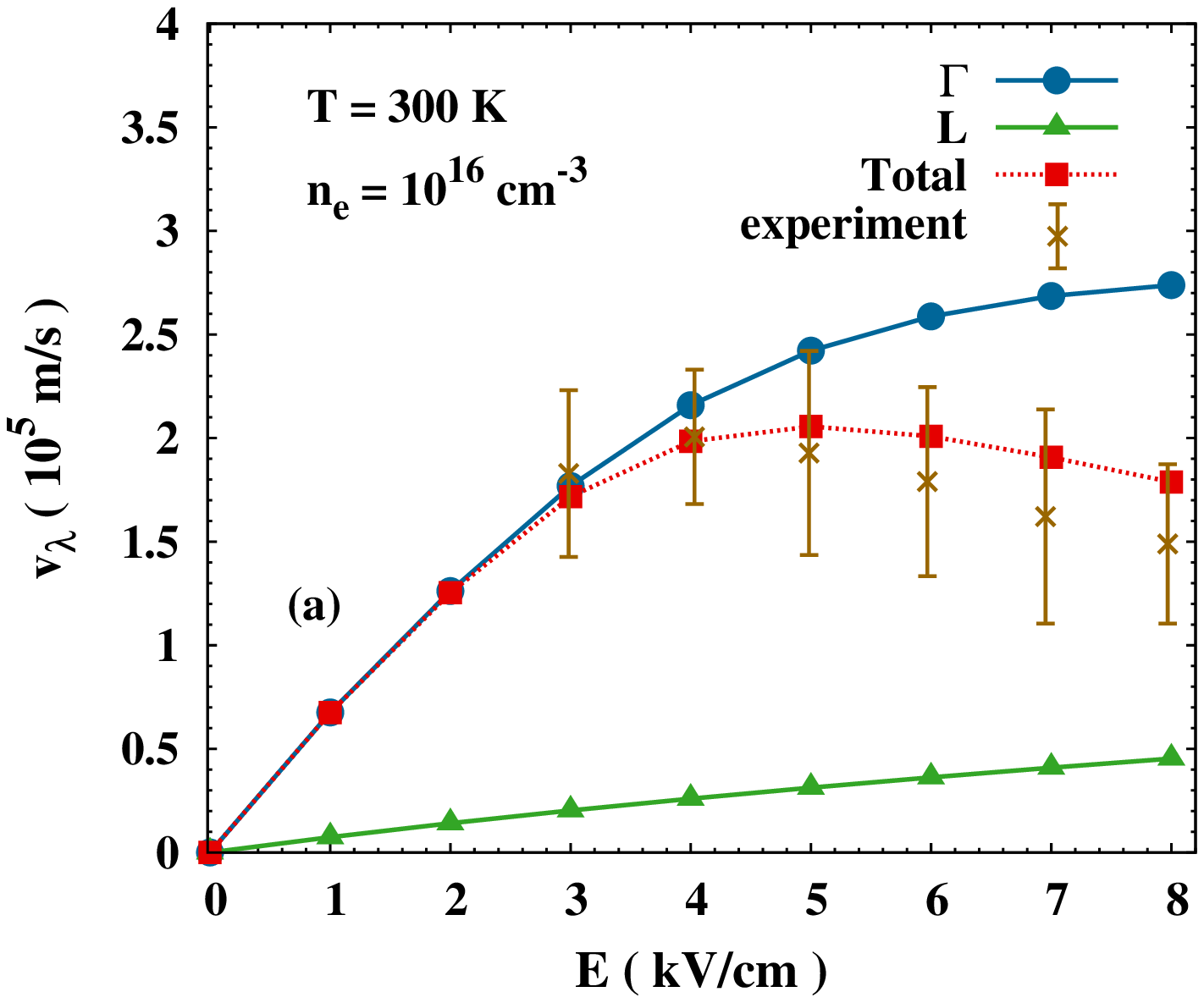}}
    \hspace{-0.7 cm}\parbox[t]{5cm}{
      \includegraphics[width=4.5cm,height=4.5 cm]{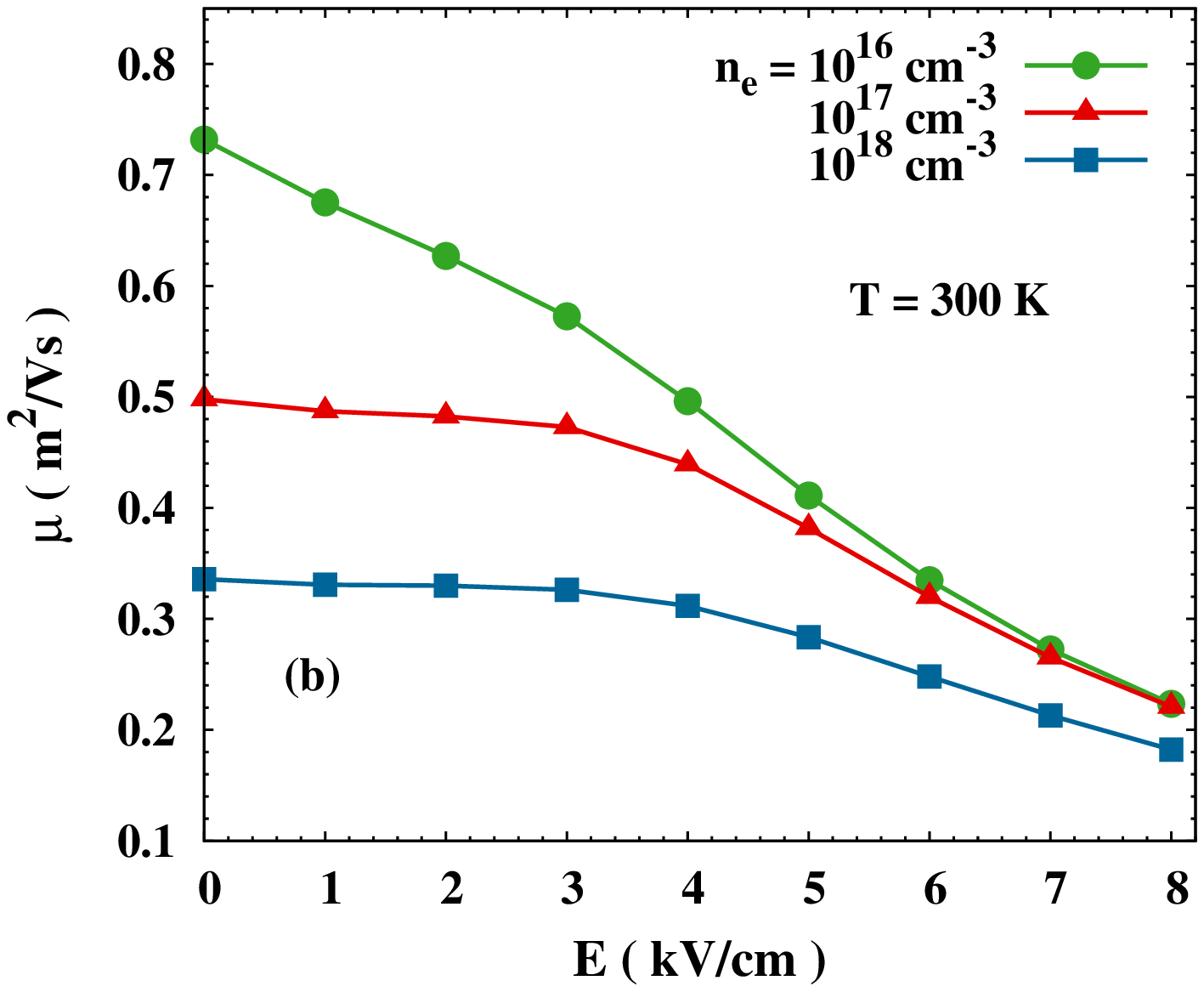}}
  \end{minipage}
  \begin{minipage}[]{10cm}
    \hspace{-1.5 cm}\parbox[t]{5cm}{
      \includegraphics[width=4.5cm,height=4.5 cm]{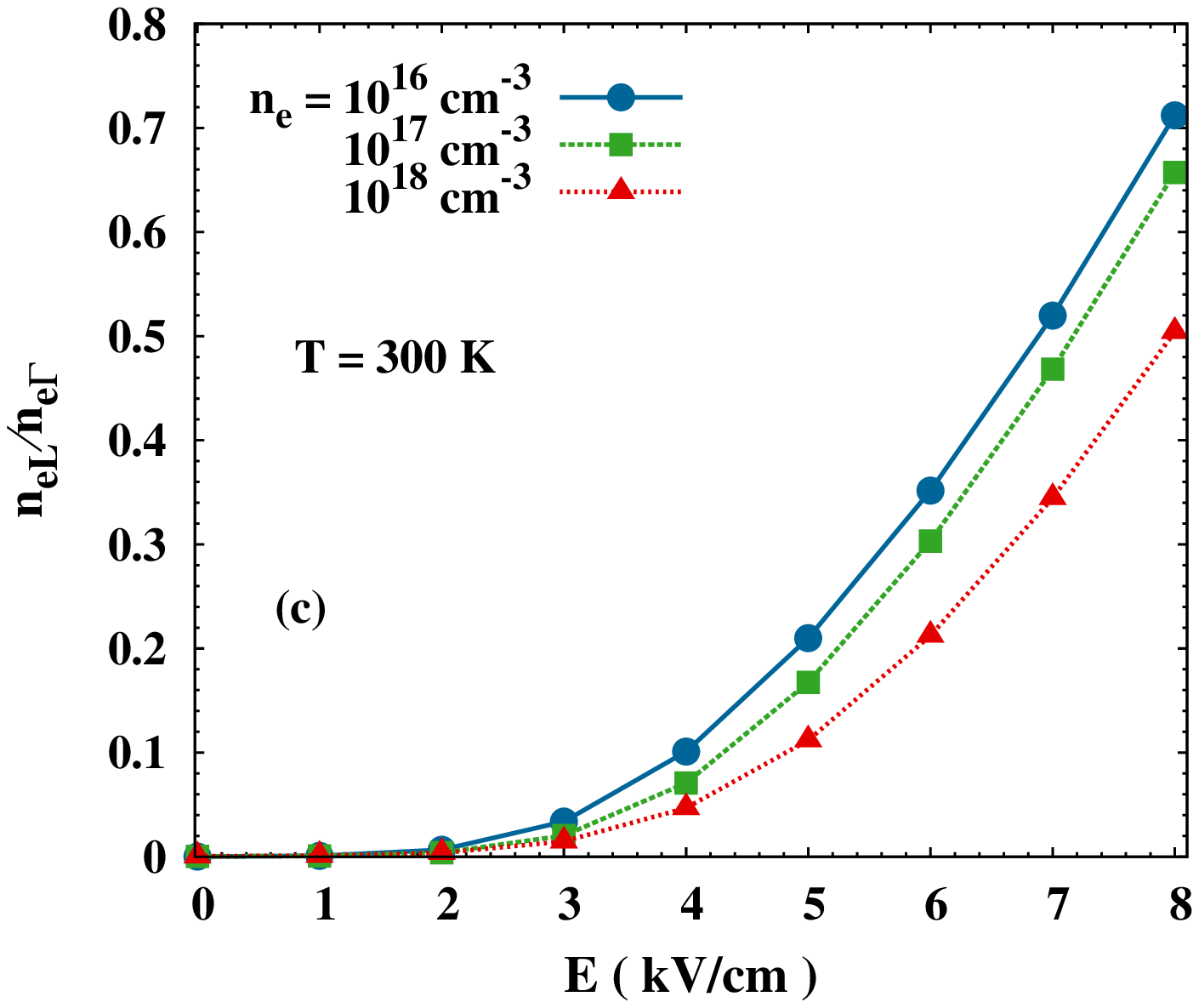}}
    \hspace{-0.7 cm}\parbox[t]{5cm}{
      \includegraphics[width=4.5cm,height=4.5 cm]{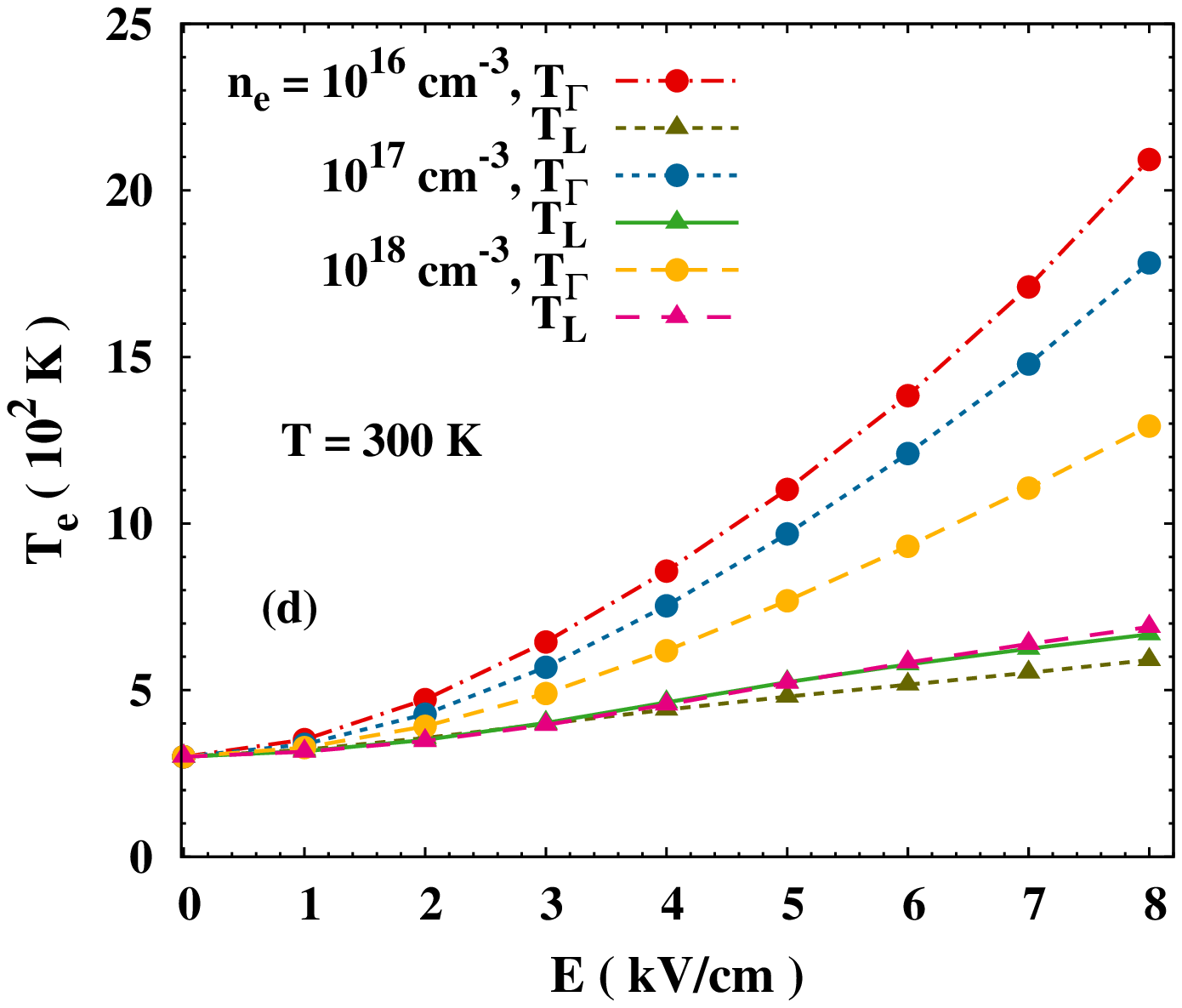}}
  \end{minipage}
  \caption{(Color online) Electric field dependences of (a) drift
    velocity $v_\lambda$ for electron density $n_e=10^{16}$~cm$^{-3}$, (b)
    mobility $\mu$, (c) ratio of electron densities  
    in the $L$ and $\Gamma$ valleys and (d) hot-electron temperature for three 
    electron densities $n_e=10^{16}$, $10^{17}$ and $10^{18}$~cm$^{-3}$. The
    lattice temperature is $T=300$~K and the 
    experimental data are taken from Ref.~\onlinecite{Kratzer}.}  
  \label{fig:All-E}
\end{figure}

\section{Drift velocity, Mobility, $L$-valley Occupation and Hot-electron
  Temperature under Electric Field}

In order to have an overview of the electric properties influenced by the high
electric field, the steady-state drift velocity $v_\lambda$ of each valley as
well as the total 
drift velocity are calculated by varying the electric field from $0$ to 
$8$~kV/cm. In Fig.~\ref{fig:All-E}(a), we plot the drift velocity $v_\lambda$
as function of electric field with
$n_e=10^{16}$~cm$^{-3}$ and $T=300$~K. The negative differential electric
conductance can be seen from the field dependence of the total drift velocity
and good agreement is reached with the experimental results.\cite{Kratzer} We
also calculate the field  dependences of the mobility and the high-valley
electron population for  three different doping densities. From
Fig.~\ref{fig:All-E}(b), one finds that 
the  mobilities first decrease slowly with the increase of electric field and
then more rapidly when the electric field is increased over $E=4$~kV/cm. This
can be understood with the help of Fig.~\ref{fig:All-E}(c) where the ratio of
electron densities in the $\Gamma$ and $L$ valleys is plotted against the electric
field $E$. It is seen that the population of
electrons in the $L$ valleys is negligible when $E < 2$~kV/cm and
approaches $10\%$ when $E\sim 4$~kV/cm. This contributes to the faster decrease
of mobilities in Fig.~\ref{fig:All-E}(b) and leads to the negative differential
electric conductance in Fig.~\ref{fig:All-E}(a). 

The hot-electron temperature $T_e$ is obtained by fitting the calculated
steady-state electron distribution of each valley with the drifted Fermi
distribution $f(\varepsilon_{{\bf k}_\lambda,E})=[{\rm
  exp}((\varepsilon_{{\bf k}_\lambda, E}-\mu_\lambda)/T_\lambda)+1]^{-1}$. Here
$\varepsilon_{{\bf k}_\lambda, E}=({\bf k}_\lambda-{{\bf
    k}}^0_\lambda)^2/2m^*_{\lambda}$ is the energy spectrum 
shifted by the electric field with ${{\bf k}}^0_\lambda$ being the drift
momentum and $\mu_\lambda$ is a fitting parameter denoting the chemical
potential in  $\lambda$ valley. It
is found that electrons in the $\Gamma$ valley carry two temperatures, one in the
higher-energy regime that overlaps with the $L$ valleys (labeled as $T_L$) and
the other in the lower-energy regime (labeled as $T_\Gamma$). Electrons in the $L$
valleys share the {\em same} temperature as those in the higher-energy regime of the 
$\Gamma$ valley due to the rapid exchange of electrons thanks to the strong
intervalley scattering. In Fig.~\ref{fig:All-E}(d), we plot $T_\Gamma$ and
$T_L$ against the electric field  at lattice temperature $T=300$~K. It is seen
that electrons in the $\Gamma$  valley are easier to be heated due to the
smaller effective mass. Moreover,  by increasing  
the electron density from $10^{16}$ to 
$10^{18}$~cm$^{-3}$, $T_\Gamma$ is effectively reduced while $T_L$
stays almost unchanged. The underlying physics is that by increasing the electron density, the
electron-impurity scattering is enhanced.  This tends to reduce the drift velocity
in each valley and suppress the ability of electrons to 
gain energy from the electric field, and therefore reduce the hot-electron
temperature.\cite{Lei2,Conwell,Seeger,Lei3} Since the electron-impurity 
scattering is the leading scattering in the $\Gamma$ valley,\cite{Jiang1} whereas the intervalley
electron-phonon scattering is dominant in the $L$ valleys,\cite{note6-L}
the drift velocity, and hence also the hot-electron
temperature of the $L$ valleys, are less affected by the electron density
compared to those of the $\Gamma$ valley.

\section{Energy-gain and Loss Rates}
For a semiconductor system under uniform electric field, the electrons
accelerate before they are scattered and thus gain energy from the electric
field. Meanwhile due to the electron-phonon scattering, the electrons transfer 
energy to the phonon system. In steady state, the electron energy-gain rate
equals the energy-loss one. \cite{Conwell,Seeger,Lei3,Lei2}

We calculate the energy-gain and loss rates in $n$-type bulk GaAs
by including only the $\Gamma$ valley. 
The energy-gain rate $\eta$ (in unit volume here and hereafter) reads\cite{Seeger} 
\begin{equation}
\eta=en_e\mu E^2 \label{energy-in}
\end{equation}
 and the energy-loss rate $\omega$ is given by\cite{Lei2}
\begin{eqnarray}
\nonumber
 \omega&=&2\sum_{{\bf q},\chi}\Omega_{{\bf q},\chi}|M({\bf q},\chi)|^2\Pi_2({\bf
   q},\Omega_{{\bf q},\chi}+\omega_0)\\
&&\mbox{}\times\left[n(\frac{\Omega_{{\bf
       q},\chi}}{T})-n(\frac{\Omega_{{\bf q},\chi}+\omega_0}{T_e})\right] , \label{energy-out}
\end{eqnarray}
with
\begin{eqnarray}
\nonumber
\Pi_2({\bf q},\omega)&=&2\pi\sum_{\bf k}[f(\varepsilon_{\bf
  k},T_e)-f(\varepsilon_{{\bf k}+{\bf q}},T_e)]\\
&&\mbox{}\times\delta(\varepsilon_{{\bf k}+{\bf q}}-\varepsilon_{\bf k}+\omega).
\end{eqnarray}
In these equations, $\Omega_{{\bf q},\chi}$ is the phonon energy with
momentum ${\bf q}$ and mode $\chi$. Note that here we only need to take account
of the intravalley electron-LO phonon scattering with $|M_{\Gamma\Gamma,{\bf
    q}}|^2$ given the Appendix~A. $\omega_0={\bf q}\cdot {\bf v}_d$ with ${\bf
  v}_d$  denoting the drift velocity. $n(x)=1/(e^x-1)$ stands for the Bose
distribution and $f(x,T_e)=1/[e^{(x-\mu)/T_e}+1]$ is the Fermi
distribution with $T_e$ and $\mu$ being the electron temperature and the
chemical potential, respectively. 

\begin{widetext}
\begin{figure}[htb]
  \begin{minipage}[]{20cm}
    \hspace{-1.5 cm}\parbox[t]{6cm}{
      \includegraphics[width=5.5cm]{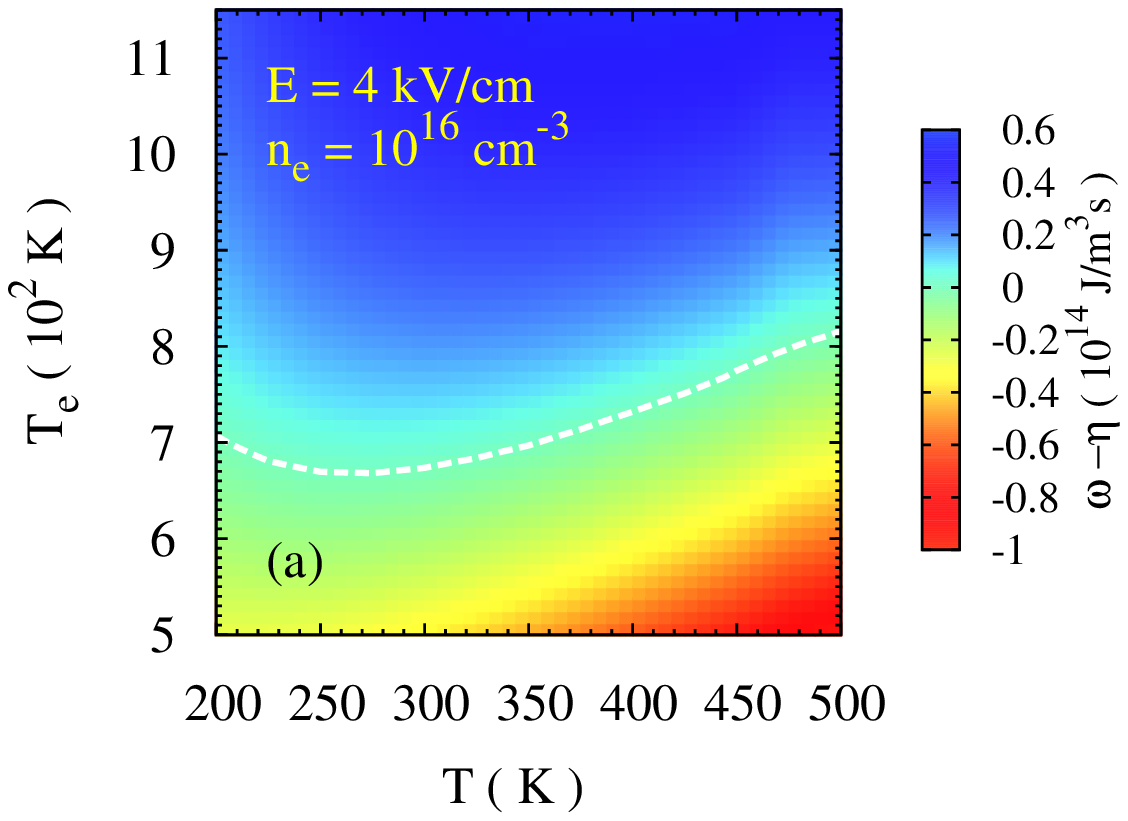}}
    \hspace{-0.7 cm}\parbox[t]{6cm}{
      \includegraphics[width=5.5cm]{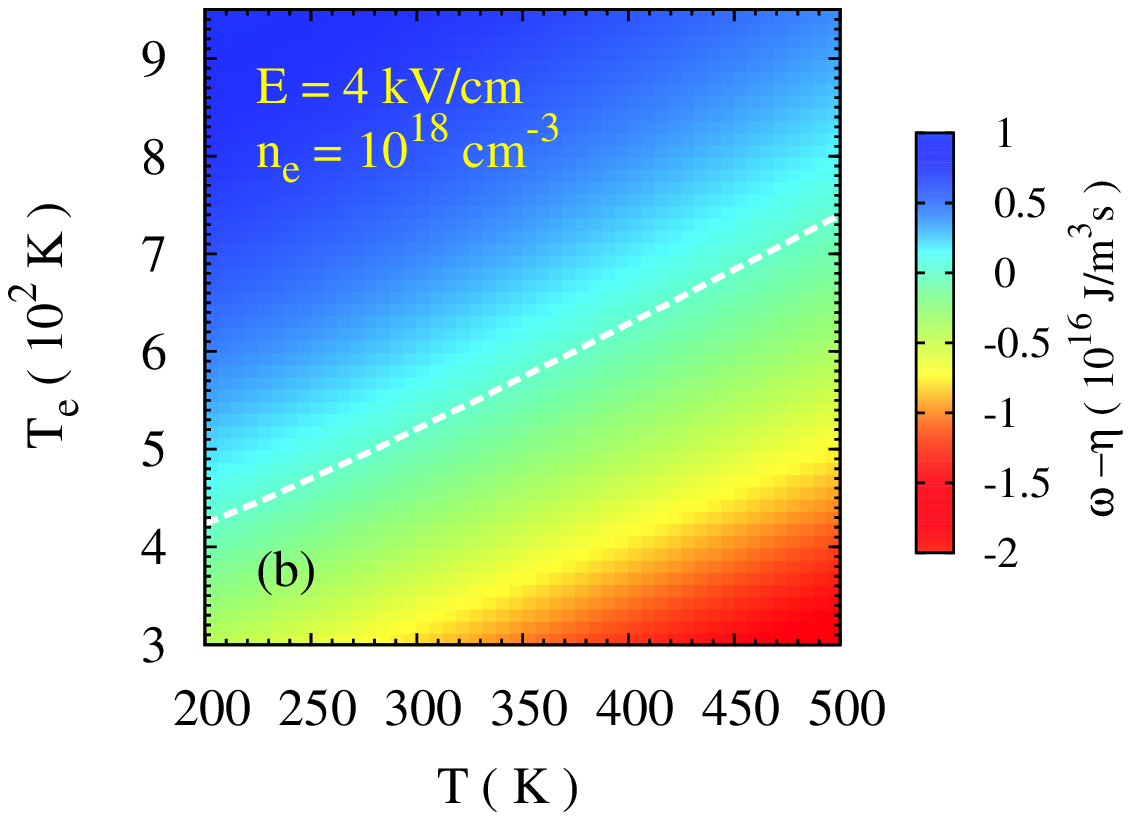}}
    \hspace{-0.7 cm}\parbox[t]{6cm}{
      \includegraphics[width=5.5cm]{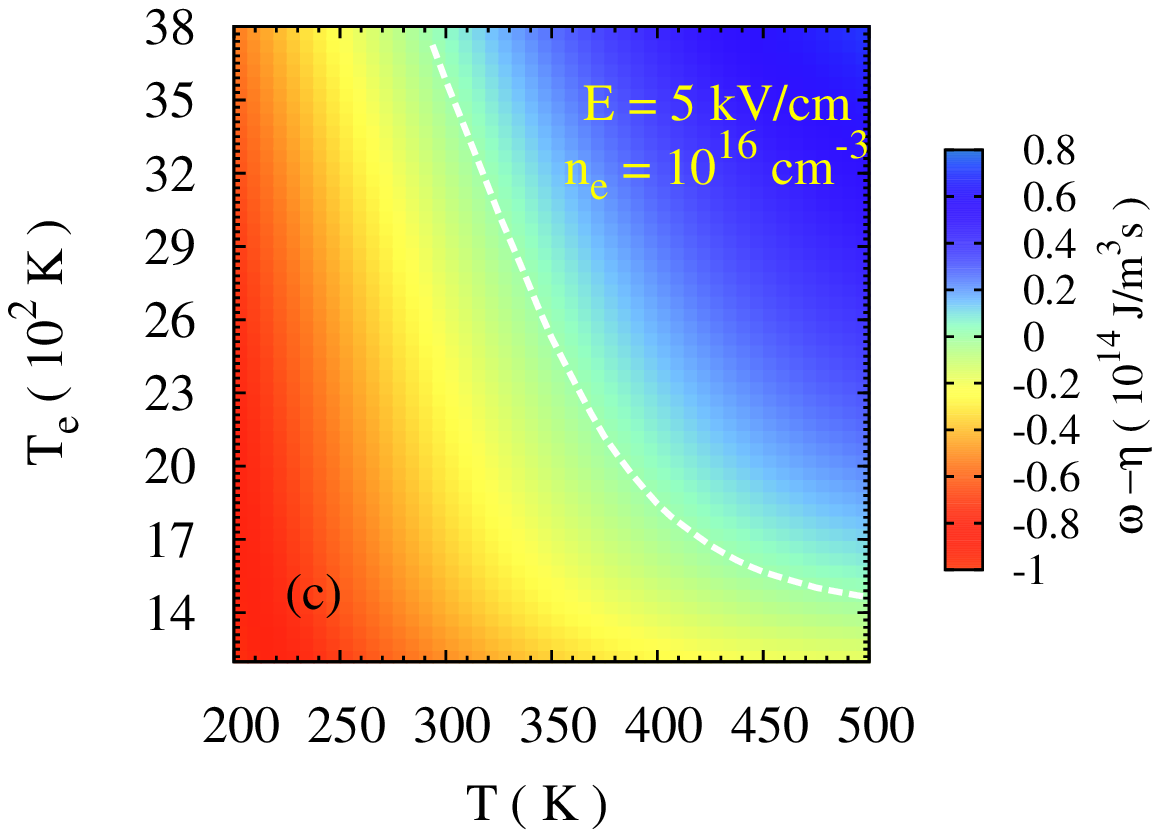}}
  \end{minipage}
  \begin{minipage}[]{17.5cm}
  \begin{center}
  \caption{(Color online) The difference of energy-gain and loss rates
    $\omega-\eta$ against the lattice temperature and hot-electron
    temperature. (a) $n_e=10^{16}$~cm$^{-3}$ and $E=4$~kV/cm; (b)
    $n_e=10^{18}$~cm$^{-3}$ and $E=4$~kV/cm; (c) $n_e=10^{16}$~cm$^{-3}$ and
    $E=5$~kV/cm. The dashed curves indicate the points where 
$\omega=\eta$, hence are just the plots of $T_e$-$T$ in equilibrium.
} 
  \label{fig:phase}
  \end{center}
  \end{minipage}
\end{figure}
\end{widetext}

From Eqs.~(\ref{energy-in}) and (\ref{energy-out}), we calculate the energy-gain
and loss rates in three cases: (i) 
$n_e=10^{16}$~cm$^{-3}$ with $E=4$~kV/cm; (ii) $n_e=10^{18}$~cm$^{-3}$ with 
$E=4$~kV/cm and (iii) $n_e=10^{16}$~cm$^{-3}$ with $E=5$~kV/cm. The lattice
temperature is varied in the range $T=200\sim 500$~K and the hot-electron temperature
$T_e$ in the proper range according to the electron density and the electric field. The
drift velocity used in the calculation are obtained by solving the KSBEs. In
Fig.~\ref{fig:phase}, we plot $\omega-\eta$ against $T$ and $T_e$ for the three
cases. The points where $\omega-\eta=0$ are indicated with dashed curve,
which is exactly the hot-electron temperature $T_e$ versus to the lattice
temperature $T$ in the steady state. By comparing the dashed curves in
Fig.~\ref{fig:phase}(a) and (b) with the corresponding curves in
Fig.~\ref{fig:t-T}(d), one notices that qualitatively, good 
agreement is reached. We note that for case (iii), due to the large electric
field, this model does not hold so well in describing the genuine system since the $L$
valleys start to play an important role and serve as the ``momentum damping
area'' where electrons are hardly drifted. Therefore in the simplified  
model, the drift effect is overestimated and the intervalley 
electron-phonon scatterings, which serve as additional energy-loss channels,
are missing. These lead to the overestimation of $T_e$. However, 
the qualitative behavior of hot-electron temperature with the lattice
temperature in the presence of extremely high electric field is still captured:
$T_e$ decreases with increasing $T$. 

The different behaviors of $T_e$ with $T$ can be understood as follows.  
The increase of the lattice temperature induces two main effects: (1) It reduces
the mobility by enhancing the electron-phonon scattering, which in turn leads to
the decrease of the energy-gain rate [see Eq.~(\ref{energy-in})]. This tends to
reduce the hot-electron temperature. (2) It also 
modulates the relative importance of phonon-emitting and absorbing processes in
the electron-phonon scattering and reduces the energy-loss rate for the
electron system (by reducing the temperature difference between the electron
and phonon systems). This tends to ``heat'' the electron system. The competing
of these two factors contributes to the complex $T$-dependence of $T_e$, hence
also that of the SRT. 
In Fig.~\ref{fig:phase}, we have shown three typical cases investigated. In
  Fig.~\ref{fig:phase}(c) with high electric field and low electron density,
the hot-electron temperature
$T_e$ is high above the lattice temperature $T$. As a result, the small increase
in $T$ does not affect the energy-loss rate much but does enhance the
scattering. This results in the leading role of the reduction of the energy-gain
rate, which in turn leads to the decrease of $T_e$. Whereas in
Fig.~\ref{fig:phase}(b) where $T_e$ is much closer to $T$ compared to that
in Fig.~\ref{fig:phase}(c), by increasing the lattice
temperature, the temperature difference between the electron
and phonon systems is effectively reduced and hence the ``heating'' effect
is more efficient. This leads to the increase of the electron temperature. 
In between, for the case with relatively low electric field 
and low electron density, a
nonmonotonic temperature dependence is expected, which is just the case in
Fig.~\ref{fig:phase}(a). 

\end{appendix}


\begin{thebibliography}{0}
\bibitem{Meier}F. Meier and B. P. Zakharchenya, {\em Optical Orientation}
  (North-Holland, Amsterdam, 1984).

\bibitem{Wolf}S. A. Wolf, D. D. Awschalom, R. A. Buhrman, J. M. Daughton, S. Von Moln\'ar,
M. L. Roukes, A. Y. Chtchelkanova, and D. M. Treger, Science {\bf 294}, 1488
(2001). 
\bibitem{Spin1}{\em Semiconductor Spintronics and Quantum Computation}, ed. by
  D. D. Awschalom, D. Loss, and N. Samarth (Springer-Verlag, Berlin, 2002).

\bibitem{Spin11}I. \v Zuti\'c, J. Fabian, and S. Das Sarma, Rev. Mod. Phys. {\bf
    76}, 323 (2004).

\bibitem{Spin12}J. Fabian, A. Matos-Abiague, C. Ertler, P. Stano, and I. \v Zuti\'c, Acta
  Phys. Slov. {\bf 57}, 565 (2007).

\bibitem{Spin13}{\em Spin Physics in Semiconductors}, ed. by M. I. D'yakonov
  (Springer, Berlin, 2008). 

\bibitem{Wu3}M. W. Wu, J. H. Jiang, and M. Q. Weng, Phys. Rep. {\bf 493}, 61
  (2010).

\bibitem{Spin2}{\em Handbook of Spin Transport and Magnetism}, ed. by E. Y. Tsymbal and
 I. \v{Z}uti\'{c} (Chapman \& Hall/CRC, 2011).


\bibitem{Seymour} R. J. Seymour, M. R. Junnarkar, and R. R. Alfano,
  Phys. Rev. B {\bf 24}, 3623 (1981).  

\bibitem{Kikkawa}J. M. Kikkawa and D. D. Awschalom, Phys. Rev. Lett. {\bf 80},
  4313  (1998).      

\bibitem{Dzhioev}R. I. Dzhioev, K. V. Kavokin, V. L. Korenev, M. V. Lazarev, B. Ya. Meltser,
  M. N. Stepanova, B. P. Zakharchenya, D. Gammon, and D. S. Katzer, Phys. Rev. B
  {\bf 66}, 245204 (2002). 

\bibitem{Buss}J. H. Bu\ss, J. Rudolph, F. Natali, F. Semond, and D. H\"aegele,
  Appl. Phys. Lett. {\bf 95}, 192107 (2009).   

\bibitem{Cundiff}M. Krau\ss, H. C. Schneider, R. Bratschitsch, Z. Chen, and
  S. T. Cundiff, Phys. Rev. B {\bf 81}, 035213 (2010).

\bibitem{Shen1}K. Shen, Chin. Phys. Lett. {\bf 26}, 067201 (2009). 

\bibitem{Bob}J. H. Bu\ss, J. Rudolph, S. Starosielec, A. Schaefer, F. Semond, Y. Cordier,
  A. D. Wieck, and D. H\"agele, Phys. Rev. B {\bf 84}, 153202 (2011).  

\bibitem{Ma1}H. Ma, Z. M. Jin, L. H. Wang, and G. H. Ma, J. Appl. Phys. {\bf
    109}, 023105 (2011). 

\bibitem{Wu4}M. W. Wu and C. Z. Ning, Phys. Stat. Sol. (b) 222, 523 (2000). 

\bibitem{Murdin} B. N. Murdin, K. Litvinenko, J. Allam, 
  C. R. Pidgeon, M. Bird, K. Morrison, T. Zhang, S. K. Clowes,
  W. R. Branford, J. Harris, and L. F. Cohen, Phys. Rev. B {\bf 72},
  085346 (2005).  

\bibitem{Oertel} S. Oertel, J. H\"ubner, and M. Oestreich,
  Appl. Phys. Lett. {\bf 93}, 132112 (2008).  

\bibitem{Jiang1}J. H. Jiang and M. W. Wu, Phys. Rev. B {\bf 79}, 125206 (2009);
   {\bf 83}, 239906(E) (2011). 


\bibitem{Zerr}K. Zerrouati, F. Fabre, G. Bacquet, G. Bandet, J. Frandon, G. Lampel and D. Paget,
  Phys. Rev. B {\bf 37} 1334 (1988).

\bibitem{Hohage} P. E. Hohage, G. Bacher, D. Reuter, and A. D. Wieck,
  Appl. Phys. Lett. {\bf 89}, 231101 (2006).     

\bibitem{Litvinenko} K. L. Litvinenko, L. Nikzad, J. Allam,
  B. N. Murdin, C. R. Pidgeon, J. J. Harris, and L. F. Cohen,
  J. Supercon. {\bf 20}, 461 (2007). 

\bibitem{Bob2}J. H. Bu\ss, J. Rudolph, F. Natali, F. Semond, and D. H\"aegele,
  Phys. Rev. B {\bf 81}, 155216 (2010).  

\bibitem{Zhu1}Y. G. Zhu, L. F. Han, L. Chen, X. H. Zhang, and J. H. Zhao,
  Appl. Phys. Lett. {\bf 97}, 262109 (2010). 

\bibitem{vanderWal}P. J. Rizo, A. Pugzlys, A. Slachter, S. Z. Denega,
  D. Reuter, A. D. Wieck, P. H. M. van Loosdrecht, and C. H. van der Wal, New
  J. Phys. {\bf 12}, 113040 (2010). 

\bibitem{Litvinenko1}K. L. Litvinenko, M. A. Leontiadou, Juerong Li, S. K. Clowes,
  M. T. Emeny, T. Ashley, C. R. Pidgeon, L. F. Cohen, and B. N. Murdin,
  Appl. Phys. Lett. {\bf 96}, 111107 (2010).

\bibitem{Barry}E. A. Barry, A. A. Kiselev, and K. W. Kim, Appl. Phys. Lett. {\bf
    82}, 3686 (2003).

\bibitem{Romer}M. R\"omer, H. Bernien, G. M\"uller, D. Schuh, J. H\"ubner, and
  M. Oestreich, Phys. Rev. B {\bf 81}, 075216 (2010).

\bibitem{Intronati}G. A. Intronati, P. I. Tamborenea, D. Weinmann, and R. A. Jalabert, arXiv:1102.4753.

\bibitem{DP1}M. I. D'yakonov and V. I. Perel', Zh. Eksp. Teor. Fiz. {\bf 60},
  1954 (1971) [Fiz. Tverd. Tela (Leningrad) {\bf 13}, 3581 (1971)];
  Sov. Phys. JETP {\bf 33}, 1053 (1971) [Sov. Phys. Solid State {\bf 13}, 3023
  (1972)].

\bibitem{Kratzer}S. Kratzer and J. Frey, J. Appl. Phys {\bf 49}, 4064 (1978).

\bibitem{Weng1}M. Q. Weng, M. W. Wu, and L. Jiang, Phys. Rev. B {\bf 69}, 245320
  (2004).
\bibitem{Weng3}M. Q. Weng and M. W. Wu, Phys. Rev. B {\bf 70}, 195318 (2004).

\bibitem{Zhang1}P. Zhang, J. Zhou, and M. W. Wu, Phys. Rev. B {\bf 77}, 235303
  (2008). 

\bibitem{Fu}J. Y. Fu, M. Q. Weng, and M. W. Wu, Physica E (Amsterdam) {\bf 40},
  2890 (2008).

\bibitem{Flatte}Y. Qi, Z. G. Yu, and M. E. Flatt\'e, Phys. Rev. Lett. {\bf 96}, 026602 (2006).

\bibitem{Wu1}M. W. Wu and C. Z. Ning, Eur. Phys. J. B {\bf 18}, 373 (2000); M. W. Wu,
  J. Phys. Soc. Jpn. {\bf 70}, 2195 (2001).

\bibitem{Wu2}M. W. Wu and H. Metiu, Phys. Rev. B {\bf 61}, 2945 (2000).

\bibitem{Vurga}I. Vurgaftman, J. R. Meyer, and L. R. Ram-Mohan,
  J. Appl. Phys. {\bf 89}, 5815 (2001).

\bibitem{Lei1}X. L. Lei, D. Y. Xing, M. Liu, C. S. Ting, and J. L. Birman,
  Phys. Rev. B {\bf 36}, 9134 (1987). 

\bibitem{Chand}N. Chand, T. Henderson, J. Klem, W. T. Masselink, R. Fischer,
  Y.-C. Chang, and H. Morko\c{c}, Phys. Rev. B {\bf 30}, 4481 (1984).

\bibitem{Dresselhaus}G. Dresselhaus, Phys. Rev. {\bf 100}, 580 (1955).

\bibitem{Ivchenko2}E. L. Ivchenko and G. E. Pikus, {\em Superlattices and Other
  Heterostructures: Symmetry and Optical Phenomena} (Springer, Berlin, 1997).

\bibitem{Korn}T. Korn, Phys. Rep. {\bf 494}, 415 (2010).

\bibitem{note2-QW}It is noted that Case (ii) is not presented because in the
  absence of the $L$ valleys, the multi-subband effect becomes important  and the
  comparison between Case (ii) and the genuine case is  meaningless. 


\bibitem{Mad}H. Ma, Z. M. Jin, G. H.  Ma, W. M. Liu, and S. H. Tang,
  Appl. Phys. Lett. {\bf 94}, 241112 (2009).
\bibitem{Jiang2}J. H. Jiang and M. W. Wu, Appl. Phys. Lett. {\bf 96}, 136101
  (2010). 

\bibitem{BYsun}B. Y. Sun, P. Zhang, and M. W. Wu, J. Appl. Phys. {\bf 108},
  093709 (2010).


\bibitem{Beschoten}B. Beschoten, E. Johnston-Halperin, D. K. Young, M. Poggio, J. E. Grimaldi,
  S. Keller, S. P. DenBaars, U. K. Mishra, E. L. Hu, and D. D. Awschalom,
  Phys. Rev. B {\bf 63}, 121202(R) (2001).  

\bibitem{Zhou1}J. Zhou, J. L. Cheng, and M. W. Wu, Phys. Rev. B {\bf 75}, 045305 (2007).

\bibitem{Leyland}W. J. H. Leyland, G. H. John, R. T. Harley, M. M. Glazov,
  E. L. Ivchenko, D. A. Ritchie, I. Farrer, 
 A. J. Shields, and M. Henini, Phys. Rev. B {\bf 75}, 165309 (2007).

\bibitem{Ruan}X. Z. Ruan, H. H. Luo, Y. Ji, Z. Y. Xu and V. Umansky,
  Phys. Rev. B {\bf 77}, 193307 (2008).  

\bibitem{MengY}L. F. Han, Y. G. Zhu, X. H. Zhang, P. H. Tan, H. Q. Ni and
  Z. C. Niu, Nanoscale Res. Lett. {\bf 6}, 84 (2011).


\bibitem{Ivchenko1}M. M. Glazov and E. L. Ivchenko,
  Pis'ma Zh. Eksp. Teor. Fiz. {\bf 75}, 476 (2002) [JETP Lett. {\bf
  75}, 403 (2002)]; Zh. Eksp. Teor. Fiz. {\bf 126}, 1465 (2004)
  [JETP {\bf 99}, 1279 (2004)].

\bibitem{Brand}M. A. Brand, A. Malinowski, O. Z. Karimov, P. A. Marsden, R. T. Harley,
  A. J. Shields, D. Sanvitto, D. A. Ritchie, and M. Y. Simmons,
  Phys. Rev. Lett. {\rm 89}, 236601 (2002).

\bibitem{Bronold}F. X. Bronold, A. Saxena, and D. L. Smith, Phys. Rev. B {\bf
    70}, 245210 (2004).


\bibitem{Weng2}M. Q. Weng and M. W. Wu, Phys. Rev. B 68, 075312 (2003).

\bibitem{Birman}J. L. Birman, M. Lax, and R. Loudon, Phys. Rev. {\bf 145}, 620
  (1966).

\bibitem{Herbert}D. C. Herbert, J. Phys. C {\bf 6}, 2788 (1973).

\bibitem{Fawcett}W. Fawcett and D. C. Herbert, J. Phys. C {\bf 7}, 1641 (1974).

\bibitem{Mick}R. Mickevi\v cius and A. Reklaitis, J. Phys.: Condens. Matter {\bf
  2}, 7883 (1990).

\bibitem{Mahan}G. D. Mahan, {\em Many-Particle Physics} (Plenum, New York, 1990).

\bibitem{Conwell}E. M. Conwell, {\em High Field Transport in Semiconductors}
  (Academic Press, New York, 1967). 

\bibitem{Lei3}X. L. Lei and C. S. Ting, Phys. Rev. B {\bf 30}, 4809 (1984).

\bibitem{Seeger}K. Seeger, {\em Semiconductor Physics: An Introduction}
  (Springer, 2004).

\bibitem{Lei2}X. L. Lei, {\em Balance Equation Approach to Electron Transport in
  Semiconductors} (World Scientific, Singapore, 2008).

\bibitem{note6-L}Compared to the genuine condition, the drift
  velocity in the $L$ valleys stays almost unchanged by setting the 
  electron-impurity scattering for $L$-valley electrons to zero but increases by more
  than $40\%$ by removing the intervalley electron-phonon scattering for both
  $E=1$ and 6~kV/cm with $n_e=10^{18}$~cm$^{-3}$ and $T=300$~K.

\end{thebibliography}
\end{document}